\documentclass[%
 reprint,
 amsmath,amssymb,
 aps,
]{revtex4-1}

\usepackage{graphicx}
\usepackage{dcolumn}
\usepackage{bm}
\usepackage{subfigure}
\usepackage{color}

\begin{document}

\preprint{APS/123-QED}

\title{Equilibrium and nonequilibrium quantum correlations \\ between two accelerated detectors}

\author{He Wang}
\affiliation{College of Physics, Jilin University, \\ Changchun 130021, China}
\affiliation{State Key Laboratory of Electroanalytical Chemistry, Changchun Institute of Applied Chemistry, \\ Changchun 130021, China}

\author{Jin Wang}
\email{jin.wang.1@stonybrook.edu}
\affiliation{Department of Chemistry and of Physics and Astronomy, Stony Brook University, Stony Brook, \\ NY 11794-3400, USA}

\date{\today}

\begin{abstract}
We quantify the quantum correlations between two accelerated detectors coupled to a scalar field in a cavity. It has been realized that an accelerated detector will experience a thermal bath, which is termed the Unruh effect. We examine the similarities and differences for quantum correlations regarding either temperature or acceleration. As the accelerations (resp.\ temperatures) of the detectors increase, the entanglement decreases to zero at some instant but the mutual information (resp.\ discord) can be amplified. As the accelerations increase but are in opposite directions, the quantum correlations decay. Importantly, we also reveal that larger quantum correlations can appear in certain nonequilibrium scenarios in which either the acceleration difference or the coupling difference becomes significant.
\end{abstract}

\maketitle


\section{\label{sec:level2} INTRODUCTION}

Quantum information science has progressed remarkably in the past three decades~\cite{r1} and promises even more revolutionary advances in the future. Quantum information is often characterized by certain quantum correlation measures such as entanglement, mutual information, discord, etc. Thought to be responsible for the unnerving action at a distance with quantum nonlocality~\cite{r2}, entanglement is a key resource in quantum computation and information, which have been studied intensively~\cite{r3}. It has been shown that entanglement has a purely quantum nature leading to advantages over classical computation. Quantum mutual information is a generalization of classical mutual information characterizing the quantum correlations between two systems~\cite{r4}, and quantum discord is the nonclassical part of the correlations characterized by the difference between quantum mutual information and classical correlation~\cite{r5}.

Most studies of quantum information and quantum correlation are focused on cases in which spacetime is flat, but in the cosmology and astrophysics of the early universe, curved spacetime is unavoidable and must be given serious consideration~\cite{r6,r7,r8,r9,r10,r11,r12}. However, quantifying quantum correlations in curved spacetime remains challenging, and how such correlations are influenced by the environment is unclear in curved spacetime. To understand quantum information and correlations in curved spacetime, we can start with the simple example of accelerated detectors (ADs). It has long been realized that the effect of local acceleration is equivalent to a thermal temperature bath, termed the Unruh or Hawking effect~\cite{r13,r14}. Therefore, it is natural to ask whether we can quantify the quantum correlations between two ADs. Are the equilibrium (same accelerations of the two detectors) quantum correlations between two ADs equivalent to those between two stationary detectors under a thermal temperature bath in flat spacetime? Moreover, what if the two accelerations or the couplings between the detectors and the vacuum field are different, and how does this nonequilibrium effect influence the quantum correlations?

Born from the marriage of special relativity and quantum mechanics, quantum field theory~\cite{r15} is suitable for studying the relativistic quantum information process. Some studies have explored the entanglement between field modes and shown that entanglement is observer-dependent and degraded from the perspective of observers in uniform acceleration because of the Unruh effect~\cite{r6,r7,r8,ref16}. With increasing research interest in the entanglement or quantum correlations between observers, particle detector models, such as the Unruh--DeWitt model, have been used to show entanglement harvesting from the coupling of the detectors to a vacuum state of a quantum field~\cite{r11,r12,r16,r17,r18,r19}. The detector usually comprises a qubit and has been shown to be useful for investigating the quantum correlations between detectors~\cite{r11,r12,r19}. There have also been approaches that involved replacing the qubit by a harmonic oscillator to serve as the detector more realistically and computing the quantum correlations between detectors through solving the Heisenberg equations of motion at different settings~\cite{r20,r21}. Furthermore, Brown \textit{et~al.}~\cite{r22} developed a method for solving for the time evolution of particle detection for ADs with a time-dependent quadratic Hamiltonian under an arbitrary number of modes in a stationary cavity. The method has the advantage of obtaining the time trajectories of an arbitrary number of detectors coupled to the vacuum field in the cavity without the need for the usual Bogoliubov transformation. By replacing a qubit (two energy levels) with a harmonic oscillator (infinitely many energy levels) in the detector model, one gains significant advantages over the standard Unruh--DeWitt (qubit-based) detector, including practical particle detection and nonperturbative treatment~\cite{r22}. Therefore, the quantum evolution of the particle detection can be solved for nonperturbatively by using the symplectic formalism for Gaussian states and operations~\cite{r22}. The evolution of the particle detection can then be obtained by solving (in general numerically) a set of coupled first-order ordinary differential equations. This formalism is especially useful for scenarios of interest in relativistic quantum theory involving quadratic Hamiltonians. However, neither (i) the quantum correlations of two equally accelerated observers and their comparisons to the corresponding stationary detectors under the common thermal temperature bath nor (ii) the quantum correlations of two differently accelerated observers and their comparisons to the corresponding stationary detectors under two different thermal temperature baths have been fully explored.

In this study, we consider two ADs moving inside a stationary cavity in flat spacetime. The detectors themselves are often considered to be effectively surrounded by two thermal baths associated with the Unruch temperature $T=a/2\pi$, where $a$ is the proper acceleration of the detector~\cite{r13}. We found that the dynamics of the quantum correlations between two equally accelerated detectors in a vacuum or two stationary detectors in a thermal field with temperature $T$ have similar qualitative trends but differ quantitatively. This shows the similarities and differences between the two detectors under equal accelerations and the same temperature for the effects on the quantum correlations. As the acceleration (temperature) of the detectors increases, the entanglement gradually disappears, while the mutual information and the quantum discord are amplified as suggested before~\cite{r23}. This shows that the quantum mutual information and discord are more robust to environmental (field) temperature fluctuations than is the entanglement. As the acceleration increases when the equal accelerations are in the opposite direction, the quantum correlations decay. This shows that the distance effect dominates the evolution.

Meanwhile, we study the quantum correlations in the nonequilibrium scenario, which herein represents the transient dynamics away from equilibrium. This nonequilibrium is influenced by the imbalanced couplings of the two equally accelerated detectors to the field in the vacuum state. We show that the nonequilibrium in time can enhance the quantum correlations. Another nonequilibrium scenario that we consider is when the two detectors are under different accelerations, mimicking situations involving different effective local temperatures. This nonequilibrium scenario persists even when the system relaxes to a steady state at long times, in contrast to the scenario in which the nonequilibrium disappears at long times in the steady state in the previous case. We refer to this as intrinsic nonequilibrium because it can be characterized by the degree of detailed balance-breaking when the system relaxes to the steady state. The acceleration difference leads to both nonequilibrium and distance effects on the quantum correlations between the two detectors.

This paper is organized as follows. In Sec.~\ref{sec:level3}, we review the continuous-variable techniques introduced in Ref.~\citenum{r22}, which hold for an arbitrary time-dependent quadratic Hamiltonian and an arbitrary number of modes in a stationary cavity. In Sec.~\ref{sec:level4}, we determine the evolution of the system, based on which we also quantify the correlation measures (the logarithmic negativity as a measure of the entanglement, the mutual information, and the Gaussian quantum discord) between the two detectors. In Sec.~\ref{sec:Equilibrium}, we consider different scenarios for the two detectors: both stationary in the thermal field or both accelerated in either the same or opposite directions. In Sec.~\ref{sec:Nonequilibrium}, we consider nonequilibrium detectors with different accelerations or different couplings to the field vacuum under the same acceleration; the stationary detectors couple to the thermal field and the ADs couple to the field vacuum, and we provide detailed discussions. Finally, we draw conclusions in Sec.~\ref{sec:level5}.

\section{\label{sec:level3} MODEL}

\subsection{Time evolution of operators}

Let us consider the global Hamiltonian and discuss how the dynamics evolve with respect to the proper time of an arbitrary detector. Throughout this paper, we use the conventions $\hbar=c=k_{B}=1$. The basic unit is the electronvolt [eV], with length and time in units of eV$^{-1}$ and temperature and mass in units of eV. Actually, the basic unit can be chosen from the energy scale of the experimental subject (e.g., a harmonic oscillator herein) in the experiment. Meanwhile, we can determine the scales of other experimental parameters (e.g., length of cavity and duration time of interaction in this paper). In all the figures, the axes are all in the corresponding units specified above. Consider a general time-dependent Hamiltonian $\hat{H}(t)$, where $t$ is the global time coordinate. The Heisenberg equation of motion of a general operator $\hat{A}(t)$ is then given by
\begin{equation}
\label{eq:1}
\frac{d\hat{A}(t)}{dt}=i[\hat{H}(t),\hat{A}(t)]+\frac{\partial\hat{A}(t)}{\partial t}.
\end{equation}

Another coordinate choice is the local proper time $\tau_{j}$, which corresponds to the world line $(t(\tau_{j}),x(\tau_{j}))$ traversed by the $j$th detector at $x$ and $t$. Applying the chain rule, the two time coordinates are associated, i.e.,
\begin{equation}\begin{split}
\label{eq:2}
\frac{d\hat{A}[t(\tau_{j})]}{d\tau_{j}}&=\frac{dt(\tau_{j})}{d\tau_{j}}\frac{d}{dt}\hat{A}(t)|_{t=t(\tau_{j})}\\
&=\frac{dt(\tau_{j})}{d\tau_{j}}(i[\hat{H}(t),\hat{A}(t)]+\frac{\partial\hat{A}(t)}{\partial t})\\
&= i[\frac{dt}{d\tau_{j}}\hat{H}(t),\hat{A}(t(\tau_{j}))]+\frac{\partial\hat{A}(t(\tau_{j}))}{\partial \tau_{j}}.
\end{split}
\end{equation}
We can rewrite the Hamiltonian as
\begin{equation}
\label{eq:3}
\hat{H}_{j}(\tau_{j})=\frac{dt}{d\tau_{j}}\hat{H}(t(\tau_{j})),
\end{equation}
and this is the Hamiltonian seen by the $j$th detector. Therefore, we can provide a table for transforming to any other time coordinate. In fact, $\frac{dt}{d\tau_{j}}$ is the redshift factor for an observer in the detector's reference frame. It provides an overall scaling of all energies in the whole system because this is what such an observer would experience. For the corresponding form in proper time, let us define
\begin{equation}
\label{eq:4}
\hat{A}_{j}(\tau_{j})=\hat{A}_{j}(t(\tau_{j})).
\end{equation}
We can use Eqs.~\eqref{eq:3} and \eqref{eq:4} to rewrite Eq.~\eqref{eq:2} for the evolution dynamics of the operator in proper time as
\begin{equation}
\label{eq:5}
\frac{d}{d\tau_{j}}\hat{A}_{j}(\tau_{j})=i[\hat{H}_{j}(\tau_{j}),\hat{A}_{j}(\tau_{j})]+\frac{\partial\hat{A}_{j}(\tau_{j})}{\partial \tau_{j}}.
\end{equation}

\subsection{\label{sec:level2} Hamiltonian of detector, field, and detector--field interactions}

We study the interactions of two detectors coupled to a quantum massless scalar field in a cavity. Let us consider an $X$--$X$ coupling form of the Unruh--DeWitt Hamiltonian~\cite{r11,r12,r13}. A general form of the Unruh--DeWitt detector field interaction is given by
\begin{equation}
\label{eq:6}
\hat{H}_{1}=\lambda(\tau)\hat{\mu}\hat{\phi}[x(\tau_{j})],
\end{equation}
where $\lambda(\tau)$ is the switching function or coupling, $\hat{\mu}$ is the monopole moment of the detector, and $\hat{\phi}[x(\tau_{j})]$ is the scalar field operator evolving along the worldline of the detector parameterized in terms of the proper time $\tau$. Here, subscript $``1"$ of the Hamiltonian represents the interaction term and subscript $``0"$ represents the free Hamiltonian term. To derive the correct form of the Hamiltonian in the Heisenberg picture, let us first write the field and monopole operators in the Schr\"{o}dinger picture. For clarity, we first consider just one detector undergoing general motion in flat spacetime with an associated proper time $\tau$ with a scalar quantum field expanded in terms of the plane-wave solutions, i.e.,
\begin{equation}
\label{eq:7}
\hat{\phi}^{S}[x(\tau)]=\sum_{n}\hat{a}_{n}v_{n}[x(\tau)]+\hat{a}_{n}^{\dagger}v_{n}[x(\tau)],
\end{equation}
where superscript $``S"$ denotes the Schr\"{o}dinger picture and $``H"$ denotes the Heisenberg picture, $\hat{a}_{n}$ and $\hat{a}_{n}^{\dagger}$ are the annihilation and creation operators, respectively, for the $n$th field mode, and $v_{n}$ is the spatial part of the $n$th mode solution to the field equations, i.e., the $n$th mode function. We consider the field and the detectors as being in a stationary cavity in flat spacetime. The cavity introduced here is for the vacuum field and the associated infrared (IR) and ultraviolet (UV) mode cutoffs. To determine these mode functions, we must impose boundary conditions for the cavity, and different boundary conditions offer different advantages, e.g.,
\begin{equation}\begin{split}
\label{eq:8}
u_n(x,t)&=\frac{1}{\sqrt{k_{n}L}}\exp(-i\omega_{n}t)\sin[k_{n}x]\ \text{or}\\
u_n(x,t)&=\frac{1}{\sqrt{k_{n}L}}\exp(-i\omega_{n}t)\exp[ik_{n}x],
\end{split}
\end{equation}
which are from the reflecting boundary condition or the periodic boundary condition, respectively. Here, $k_{n}$ is the wave vector for the $n$th field mode. For the massless scalar field, we have $\omega_{n}=|k_{n}|$, with $k_{n}=n\pi/L$ under the reflecting boundary condition or $k_{n}=2n\pi/L$ under the periodic boundary condition, where $L$ is the length of the cavity. The periodic boundary condition is preferred herein because it avoids the blueshift-induced decoupling~\cite{r22}.

In the Schr\"{o}dinger picture, the monopole moment $\hat{\mu}$ of the detector is given as
\begin{equation}
\label{eq:9}
\hat{\mu}^{S}=\hat{a}_{d}+\hat{a}_{d}^{\dagger},
\end{equation}
where $\hat{a}_{d}$ and $\hat{a}_{d}^{\dagger}$ are the annihilation and creation operators, respectively, for the detector. Now, we write the whole Hamiltonian $\hat{H}^{S}$ in the Schr\"{o}dinger picture with respect to the detector's proper time under the transformation table as
\begin{equation}\begin{split}
\label{eq:10}
\hat{H}^{S}&=\hat{H}_{0}^{S}+\hat{H}_{1}^{S}\\
&=\frac{dt(\tau)}{d\tau}\sum_{n}\omega_{n}\hat{a}_{n}^{\dagger}\hat{a}_{n}+\Omega\hat{a}_{d}^{\dagger}\hat{a}_{d}\\
&+\lambda(\tau)(\hat{a}_{d}+\hat{a}_{d}^{\dagger})\sum_{n}(\hat{a}_{n}v_{n}[x(\tau)]+\hat{a}_{n}^{\dagger}v_{n}[x(\tau)]),
\end{split}
\end{equation}
where $\Omega$ is the characteristic frequency of the detector. The first terms of Eq.~\eqref{eq:10} represent the Hamiltonian of the scalar field, the second term represents the Hamiltonian of the detector, and the third terms represent the interactions between the field (modes) and the detector. Notice that the form of the complete Hamiltonian in the Heisenberg picture coincides with the form of the Hamiltonian in the Schr\"{o}dinger picture.

It is straightforward to write the most general $X$--$X$-type Hamiltonian for an arbitrary number of detectors moving on general trajectories with different proper times $\tau_{j}$ and with time-dependent couplings, i.e.,
\begin{equation}\begin{split}
\label{eq:11}
\hat{H}^{H}&=\hat{H}_{0}^{H}+\hat{H}_{1}^{H}\\
&=\sum_{n=1}^{N}\omega_{n}\hat{a}_{n}^{\dagger}\hat{a}_{n}+\sum_{j=1}^{M}\frac{d\tau_{j}}{dt}[\Omega\hat{a}_{d_{j}}^{\dagger}\hat{a}_{d_{j}}\\
&+\sum_{n=1}^{N}\lambda_{nj}(t)(\hat{a}_{d_{j}}+\hat{a}_{d_{j}}^{\dagger})(\hat{a}_{n}v_{n}[x(\tau)]+\hat{a}_{n}^{\dagger}v_{n}[x(\tau)])],
\end{split}
\end{equation}
where the first terms represent the Hamiltonian of the scalar field, the second terms represent the Hamiltonian of the detectors, and the third terms represent the interactions between the field (modes) and the detectors. From Eq.~\eqref{eq:11}, our cavity setting with length $L$ introduces a natural IR cutoff of the field modes, and a $UV$ cutoff is introduced for the sake of the computation and the resolution of the detectors.

\subsection{\label{sec:level2} Evolution dynamics of detectors}

Conventionally, there have been two main approaches to quantum information processing~\cite{r24}. On one hand, quantum information is encoded into systems with a discrete and finite number of degrees of freedom, such as qubits; typical examples of qubit implementations are the nuclear spins of individual atoms in a molecule, the polarization of photons, and the ground/excited states of trapped ions. On the other hand, a continuous approach has also been investigated, with the correlations encoded in the degrees of freedom through continuous variables; typical examples of the continuous-variable implementations are the position and momentum of a particle. In the present study, we mainly use the second approach. Specifically, both the detectors and the scalar field modes are described as continuous variables to account for the underlying bosonic degrees of freedom. We use the Hamiltonian of Eq.~\eqref{eq:11} in the Heisenberg picture to evolve the quadrature operators $(\hat{q}_{d_{j}}(t),\hat{p}_{d_{j}}(t))$ for each detector and $(\hat{q}_{n}(t),\hat{p}_{n}(t))$ for each field mode, where $j$ runs over all detectors and $n$ runs over all field modes. These coordinate and momentum operators satisfy the standard canonical commutation relations as follows: $[\hat{q}_{d_{i}}(t),\hat{p}_{d_{j}}(t)]=i\delta_{ij}$, $[\hat{q}_{n}(t),\hat{p}_{k}(t)]=i\delta_{nk}$. We organize these operators in a phase-space vector form as
\begin{equation}
\label{eq:12}
\hat{X}=(\hat{q}_{d_{1}},\hat{p}_{d_{1}},...\hat{q}_{d_{j}},\hat{p}_{d_{j}}...\hat{q}_{1},\hat{p}_{1},...\hat{q}_{n},\hat{p}_{n}).
\end{equation}
The vector $\hat{X}$ in a phase space naturally has a symplectic form $[\hat{X},\hat{X}^{T}]=i\Delta=i\oplus_{i}\begin{bmatrix} 0 & 1 \\ -1 & 0 \end{bmatrix}$, where $i$ runs over all degrees of freedom (both detectors and field modes). We consider the state of our detector--field system to be a Gaussian state~\cite{r25} for a continuous-variable ensemble. A Gaussian state is defined as any state whose characteristic functions and quasi-probability distributions are Gaussian functions in the quantum phase space. In general, a Gaussian state is fully characterized by its first and second canonical moments only~\cite{r24}. In our study, a Hamiltonian that is quadratic in the operators with no linear terms as in Eq.~\eqref{eq:11} is considered. Consequently, all states are Gaussian with no displacement, and all evolutions are homogeneous Gaussian unitary ones. Therefore, we may neglect the first moments, and the state is fully described by the second moments or the covariance matrix $\sigma$, i.e.,
\begin{equation}
\label{eq:13}
\sigma_{ij}=\left\langle\hat{x}_{i}\hat{x}_{j}+\hat{x}_{j}\hat{x}_{i}\right\rangle-2\left\langle\hat{x}_{i}\right\rangle\left\langle\hat{x}_{j}\right\rangle.
\end{equation}

The time evolution of the entire covariance matrix, including both the detector and the field, is governed by the equation of unitary evolution
\begin{equation}
\label{eq:14}
\sigma(t)=S(t)\sigma(0)S^{T}(t),
\end{equation}
where $S$ is a symplectic matrix: $S\Delta S^{T}=S^{T}\Delta S$ and $\hat{X}(t)=S(t)\hat{X}(0)$. In Hilbert space, the evolution of a state depends on the time evolution operator $\hat{U}$, with $S$ playing the same role in the quantum phase space. A general time-dependent quadratic Heisenberg-picture Hamiltonian can be written as
\begin{equation}
\label{eq:15}
\hat{H}=\hat{X}^{T}F(t)\hat{X},
\end{equation}
where $F(t)$ is a Hermitian matrix of c-numbers determined only by the explicit time-dependence of the Hamiltonian; for more details about deriving $F(t)$, see Appendix~\ref{sec:Appendix:A}. The Heisenberg equation for the time evolution of the quadratures is given by
\begin{equation}
\label{eq:16}
\frac{d\hat{X}}{dt}=i[\hat{H},\hat{X}].
\end{equation}
Using Eqs.~\eqref{eq:15} and \eqref{eq:16} and the definition of $\Delta$, one derives
\begin{equation}
\label{eq:17}
\frac{d\hat{X}}{dt}=\Delta F^{sys}\hat{X},
\end{equation}
where $F^{sys}=F+F^{T}$. This is just the quantum phase-space version of the Schr\"{o}dinger equation. Using $\hat{X}(t)=S(t)\hat{X}(0)$ and $[\hat{X},\hat{X}^{T}]=i\Delta$, finally we obtain
\begin{equation}
\label{eq:18}
\frac{dS}{dt}=\Delta F^{sys}S(t).
\end{equation}

Note that once we solve Eq.~\eqref{eq:18} with the initial condition $S(0)=I_{2d_{n}+2N}$ such that $\hat{X}(0)=S(0)\hat{X}(0)$, we know the evolution of the detector--field system (the subscripts $d_{n}$ and $N$ mean that there are $d_{n}$ detectors and $N$ field modes). In the particularly simple cases in which the Hamiltonian at different times commutes, the solution of Eq.~\eqref{eq:18} can be obtained analytically and is simply $S(t)=\exp[\Delta F^{sys} t]$. From Eq.~\eqref{eq:11}, our cavity setting with length $L$ introduces a natural IR cutoff of the field modes, and a UV cutoff is introduced for the sake of the computation and the resolution of the detectors. Using causality analysis in Appendix~\ref{sec:Appendix:B}, we provide an effective criterion for choosing a reasonable cutoff with required arbitrary high precision.

\section{\label{sec:level4} EVOLUTION OF SYSTEM AND MEASURES OF QUANTUM CORRELATIONS}

\subsection{\label{sec:Covariance and S matrix evolution} Covariance and S matrix evolution}

In what follows, we consider two ADs moving inside a stationary cavity in flat spacetime. The two separable detectors are initially stationary inside the cavity. They comprise a harmonic oscillator rather than a qubit, and they both have the same characteristic frequency $\Omega$. We focus on one-dimensional dynamics, so we choose a smooth cavity of length $L$. Therefore, there is a natural IR cutoff for the field modes, and a UV cutoff is introduced for the sake of the computation and the resolution of the detectors. The Hamiltonian is
\begin{equation}\begin{split}
\label{eq:19}
&\hat{H}^{accelerated}=\sum_{n=1}^{N}\omega_{n}\hat{a}_{n}^{\dagger}\hat{a}_{n}+\sum_{j=1}^{2}\frac{d\tau_{j}}{dt}[\Omega\hat{a}_{d_{j}}^{\dagger}\hat{a}_{d_{j}}\\
&+\sum_{n=1}^{N}\lambda_{nj}(t)(\hat{a}_{d_{j}}+\hat{a}_{d_{j}}^{\dagger})(\hat{a}_{n}v_{n}[x(\tau)]+\hat{a}_{n}^{\dagger}v_{n}[x(\tau)])],
\end{split}
\end{equation}
\begin{equation}\begin{split}
\label{eq:20}
&\hat{H}^{stationary}=\sum_{n=1}^{N}\omega_{n}\hat{a}_{n}^{\dagger}\hat{a}_{n}+\sum_{j=1}^{2}[\Omega\hat{a}_{d_{j}}^{\dagger}\hat{a}_{d_{j}}\\
&+\sum_{n=1}^{N}\lambda_{nj}(t)(\hat{a}_{d_{j}}+\hat{a}_{d_{j}}^{\dagger})(\hat{a}_{n}v_{n}[x(\tau)]+\hat{a}_{n}^{\dagger}v_{n}[x(\tau)])],
\end{split}
\end{equation}
from which $F^{sys}$ can be derived. From Eq.~\eqref{eq:18}, we can calculate the evolution of the symplectic matrix $S(t)$.

An AD will experience a thermal bath in which the temperature is associated with $T=a/2\pi$, where $a$ is the proper acceleration of the detector. The initial continuous variables for the covariances are
\begin{equation}
\label{eq:21}
\sigma_{f}^{0}=I_{2N}
\end{equation}
or
\begin{equation}
\label{eq:22}
\sigma_{f}^{0}=\oplus_{i}\begin{bmatrix} v_{i} & 0 \\ 0 & v_{i} \end{bmatrix}.
\end{equation}
Equation~\eqref{eq:21} denotes a field in the vacuum state ($I_{2N}$ is an identity matrix), and Eq.~\eqref{eq:22} denotes a field in the thermal state, where $v_{i}=\frac{\exp(\frac{\omega_{i}}{T})+1}{\exp(\frac{\omega_{i}}{T})-1}$, and $T$ is the temperature of the field~\cite{r24}. For both accelerated and stationary detectors, we have
\begin{equation}
\label{eq:23}
\sigma_{d}^{0}=I_{4},
\end{equation}
and the initial continuous covariance matrix of the whole system is given as
\begin{equation}
\label{eq:24}
\sigma(0)=\sigma_{d}^{0}\oplus\sigma_{f}^{0}.
\end{equation}

Once determined from Eq.~\eqref{eq:14}, the covariance-matrix evolution of our system can be obtained and takes the generic form
\begin{equation}
\label{eq:25}
\sigma(t)=\begin{bmatrix} \sigma_{d} & \gamma \\ \gamma^{T} & \sigma_{f} \end{bmatrix},
\end{equation}
where $\sigma_{d}$ and $\sigma_{f}$ represent the $4M\times4M$ and $4N\times4N$ covariance matrices of the reduced states of the $M$ detectors and $N$ field modes, respectively. The matrix $\gamma$ contains the information about the correlations between the detectors and the field. We can compute the correlation measures between the detectors, such as the logarithmic negativity for entanglement, the mutual information, and the quantum discord. The covariance matrix of the detector--detector state obtained upon evolution is in the generic form (only considering the two-detector case here)
\begin{equation}
\label{eq:26}
\sigma(t)=\begin{bmatrix} \sigma_{d_{1}} & \gamma_{12} \\ \gamma_{12}^{T} & \sigma_{d_{2}} \end{bmatrix},
\end{equation}
where $\sigma_{d_{1}}$ and $\sigma_{d_{2}}$ describe the reduced states of detectors $1$ and $2$, respectively. The matrix $\gamma_{12}$ stores information about the correlations between the two detectors; for example, the detectors are uncorrelated (i.e., in a product state) if and only if all entries of $\gamma_{12}$ are zero.

\subsection{\label{sec:Quantum correlation measures} Quantum correlation measures: entanglement, mutual information, and discord}

The von~Neumann entropy of a general Gaussian state is given by~\cite{r23}
\begin{equation}
\label{eq:27}
S_{VN}=(\sigma)=\sum_{i=1}^{N}s(v_{i}),
\end{equation}
\begin{equation}
\label{eq:28}
s(x)=\frac{x+1}{2}\log_{2}(\frac{x+1}{2})-\frac{x-1}{2}\log_{2}(\frac{x-1}{2}),
\end{equation}
where $v_{i}$ is the $i$th symplectic eigenvalue (i.e., the orthogonal eigenvalues of the matrix $|i\Delta\sigma|)$. The mutual information between the detectors, which quantifies the total amount of the correlation between the subsystems that can potentially be useful for the computational tasks, then follows the usual form~\cite{r6}
\begin{equation}
\label{eq:29}
I_{MI}=S_{VN}(\sigma_{d_{1}})+S_{VN}(\sigma_{d_{2}})-S_{VN}(\sigma_{d}).
\end{equation}

It is easy to see that when $\gamma_{12}=0$, $I=0$, in which case there is no correlation between the two detectors. Defining the quantities $\alpha=Det[\sigma_{d_{1}}]$, $\beta=Det[\sigma_{d_{2}}]$, $\gamma=Det[\gamma_{12}]$, and $\delta=Det[\sigma_{d}]$, the logarithmic negativity for characterizing the entanglement between the detectors is given as~\cite{r23}
\begin{equation}
\label{eq:30}
E_{N}=Max[0,-\log_{2}(\tilde{v}_{-})],
\end{equation}
where $\tilde{v}_{-}$ is the smaller of the state's partially transposed symplectic eigenvalues and is given by $\tilde{v}_{-}=\sqrt{\frac{\tilde{\Delta}-\sqrt{\tilde{\Delta}^{2}-4\delta}}{2}}$, where $\tilde{\Delta}=\alpha+\beta-2\gamma$.

The quantum discord is a measure of the pure quantum part of the correlations obtained by subtracting the classical correlations from the mutual information. It is a measure of the information that cannot be extracted without joint measurements. The discord is a good indicator of the quantum nature of quantum correlation~\cite{r5} and is given by~\cite{r26}
\begin{equation}
\label{eq:31}
D(1:2)=s(\sqrt{\beta})-s(\sqrt{v_{1}})-s(\sqrt{v_{2}})+s(\sqrt{E}),
\end{equation}
where $v_{1}$ and $v_{2}$ are the symplectic eigenvalues of $\sigma_{d}$ and $E$ is defined as
\begin{equation}
\label{eq:30}
E=\left\{
\begin{aligned}
&\frac{2\gamma^{2}+(-1+\beta)(-\alpha+\beta)}{(-1+\beta)^{2}}\\
&+\frac{2|\gamma|\sqrt{\gamma^{2}+(-1+\beta)(-\alpha+\beta)}}{(-1+\beta)^{2}}\\
&\text{if}\ (-\alpha\beta+\delta)^{2}\le(-1+\beta)(\alpha+\delta)\gamma^{2}, \text{or}\\
&\frac{\alpha\beta-\gamma^{2}+\delta}{2\beta}\\
&-\frac{\sqrt{\gamma^{4}+(-\alpha\beta+\delta)^{2}-2\gamma^{2}(\alpha\beta+\delta)}}{2\beta}\\
& \text{otherwise.}
\end{aligned}
\right.
\end{equation}
In general, the discord is not symmetric, i.e., $D(1:2)\neq D(2:1)$.

\section{\label{sec:Equilibrium} EQUILIBRIUM QUANTUM CORRELATIONS BETWEEN TWO EQUALLY ACCELERATED DETECTORS}

\subsection{Accelerations of detectors in same direction}

\subsubsection{Entanglement}

In what follows, we consider two detectors in the equilibrium state, which means that they are either stationary and have the same temperature or have the same accelerated motion in the cavity. The world lines are given by $(t,\pi,0,0)$ and $(t,2\pi,0,0)$ for the two stationary detectors named Alex and Robb and by $(a^{-1} \sinh(a\tau),a^{-1}(\cosh(a\tau)-1),0,0)$ and $(a^{-1} \sinh(a\tau),a^{-1}(\cosh(a\tau)-1+a\pi),0,0)$ for the two ADs named Alice and Bob. We consider two different setups: (i) the two detectors accelerate with the same acceleration starting from the vacuum state; (ii) the two separated detectors start from the vacuum state and are then immersed in a stationary manner in a thermal field. We are interested in exploring the behaviors of the entanglement and the mutual information as well as the quantum discord between the detectors from the field in different settings. In a cavity, the effect of the boundary on the detector's interaction with the quantum field becomes important, especially for an AD. In~\cite{ref27}, the authors circumvented the boundary effect by arranging a series of cavities wall to wall. In the present study, we adopt another approach that considers the interaction between the detectors and the field to be switched on far away from the edge (as was considered in~\cite{Universality28} to avoid the boundary effect). To do so, we consider a Gaussian distribution in time, i.e., $\lambda(\tau)=\lambda_{0}\exp(-\frac{(\tau-\tau_{0})^2}{2\varepsilon})$, through which setting the AD decouples from the field when the latter is close to the boundary.

\begin{figure}[htbp!] \centering
\subfigure[ ] { \label{fig:1a}
\includegraphics[width=2.0in]{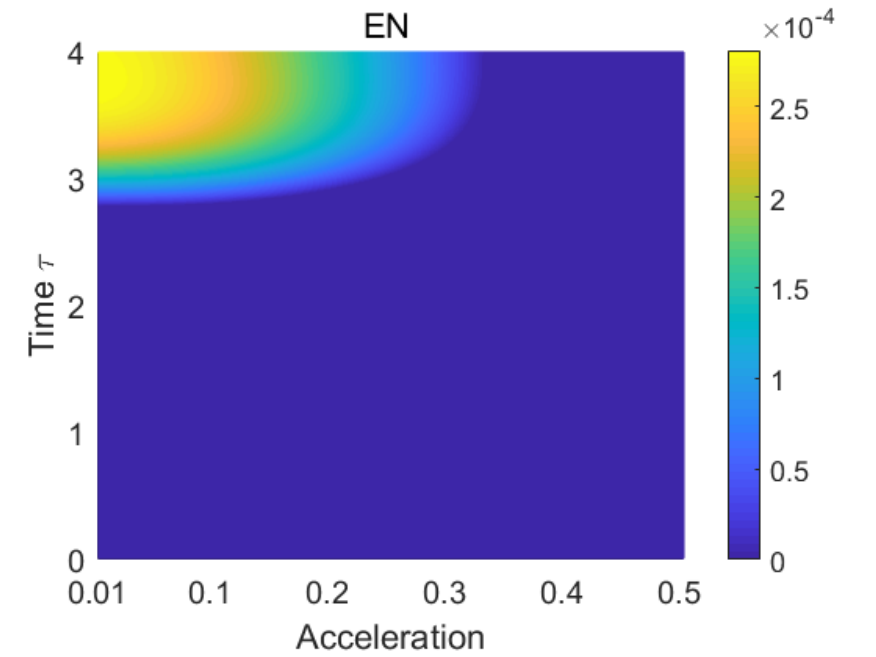}
}
\subfigure[ ] { \label{fig:1b}
\includegraphics[width=2.0in]{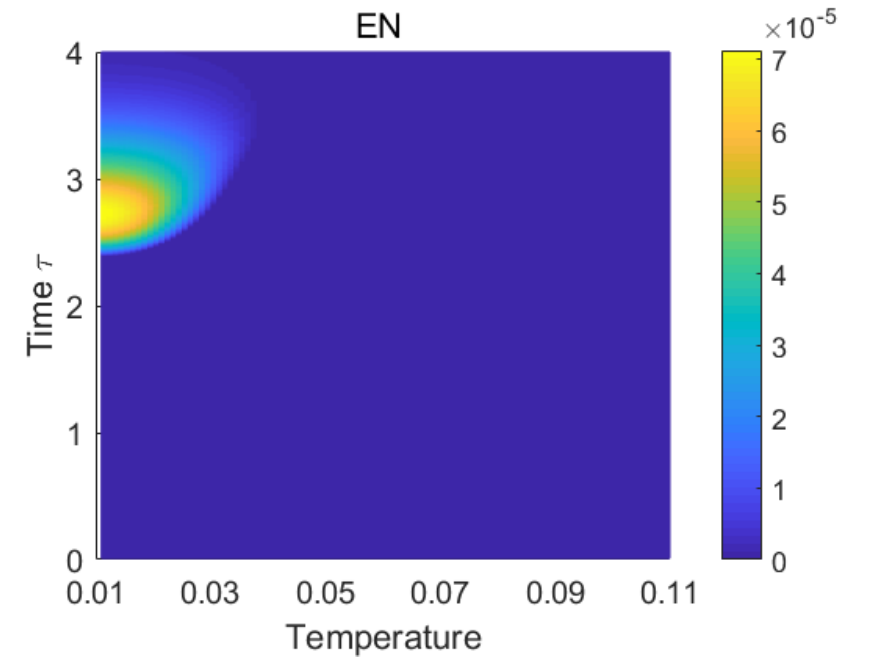}
}
\caption{(a) Variation of $E_N$ with acceleration and proper time $\tau$. The acceleration of the detectors increases from $0.01$ as they accelerate with the same proper acceleration and are always a distance $\pi$ apart. (b) Variation of $E_N$ with temperature of field and proper time $\tau$. $T$ is scaled by multiplication by $2\pi$. The temperature of the detectors increases from $0.01$ as they remain stationary at a distance $\pi$ apart. The cavity length is $4\pi$, while the detector's response frequency is set as $\Omega=3/2$, which indicates that the detector is in resonance with the third field mode. That means that only finite modes need be considered because the detector does not respond easily to higher-energy modes. The number of field modes is set as $80$, and the coupling strength for both detectors is set as $\lambda(\tau) =0.05\exp(-\frac{(\tau-1.5)^2}{2})$. }
\label{fig}
\end{figure}

The logarithmic negativity $E_{N}$ between the detectors is a measure of the entanglement. We show this entanglement measure in two cases in Figs.~\ref{fig:1a} and \ref{fig:1b}. Figure~\ref{fig:1a} shows $E_{N}$ between the two ADs varying with the acceleration and the proper time, while Fig.~\ref{fig:1b} shows $E_{N}$ between the two stationary detectors varying with the temperature of the field and the proper time (equal to the coordinate time). We choose the proper time because the period of the correlation dynamics of the ADs associated with the response frequency of the detector is a constant in the detector's proper time; in other words, we can avoid the redshift effect that appears in the coordinate time.

The cavity setting limits the plot range. Strictly speaking, we can see only part of the effect of the detectors, especially for the ADs, but nevertheless some interesting behaviors have been found. The two initially unrelated detectors can harvest the entanglement from the field, with the ADs harvesting later than the stationary detectors. Although the quantitative details differ between the two cases, the overall qualitative trends in time and in acceleration (temperature) are similar. Entanglement is resurrected from dead and dies again in time, and this behavior appears later in time for the ADs than for the stationary detectors in the thermal bath.

The fluctuating oscillation pattern in time of the entanglement might be due to the quantum recurrence (live-and-die) behavior from the nonMarkovian effect~\cite{r27}. We do not use the Born--Markov approximation in the computation. The environment is nonMarkovian, which preserves a certain memory of the system and then affects the subsequent system evolution. The nonMarkovian effect can lead to information flowing back from the environment to the system~\cite{ref28}, thus the entanglement within the system may exhibit an oscillating pattern~\cite{Ref27,Ref29} (this is easier to check when the coupling is constant). The eventual decreasing trend of the entanglement in time may come from the fact that the stronger the thermal noise, the more thermal excitations the detectors can capture. At long times, the systems reach equilibrium and become thermal mixed states lacking a quantum nature, which weakens the harvesting entanglement. Meanwhile, the thermal noise strengthens the system--field interaction and further boosts the system--field entanglement. By contrast, the detector--detector entanglement weakens because of the monogamous relationship of quantum correlations~\cite{Ref28}. Therefore, the thermal noise harms the entanglement within the system in general. Besides, the coupling constant decreases over time once $\tau > \tau_{0}$. We can then omit the memory effect of the environment, considering that there is no feedback from environment to system as long as the coupling to the environment is sufficiently weak~\cite{ref29}.

At fixed times, as the acceleration (temperature) increases, the entanglement decreases monotonically. This can be seen as a consequence of the Unruh effect: the local acceleration is equivalent to the temperature. When the acceleration or equivalently the temperature is sufficiently high, there is no entanglement in Figs.~\ref{fig:1a} and \ref{fig:1b}. As the acceleration (temperature) increases, the thermal effects become more pronounced and the decoherence becomes more significant, leading to reduced entanglement. Note that the entanglement reduction is faster for the stationary detectors in a thermal bath than for the ADs.

\subsubsection{Quantum mutual information and quantum discord}

\begin{figure}[htbp!] \centering
\subfigure[ ] { \label{fig:2a}
\includegraphics[width=2.0in]{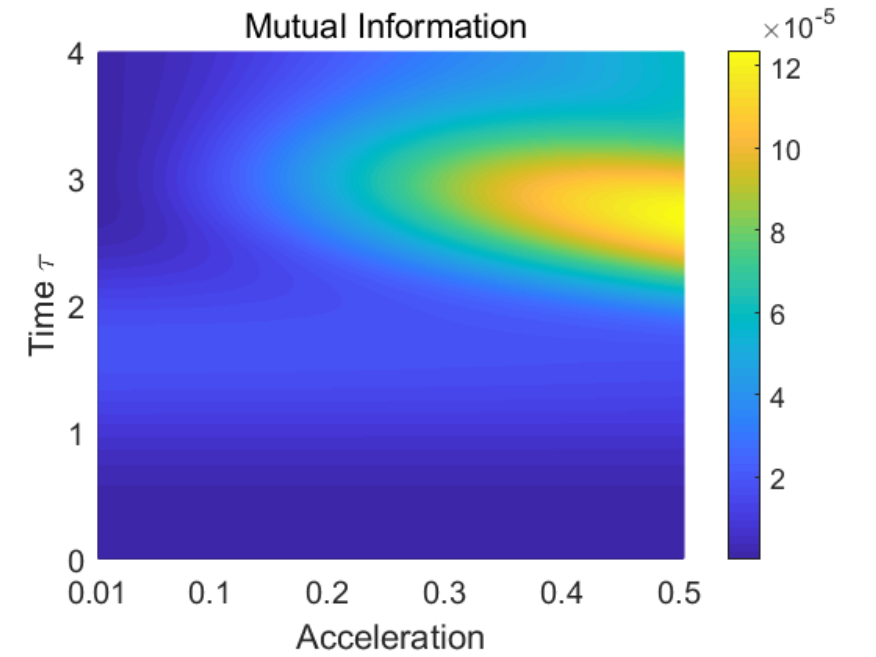}
}
\subfigure[ ] { \label{fig:2b}
\includegraphics[width=2.0in]{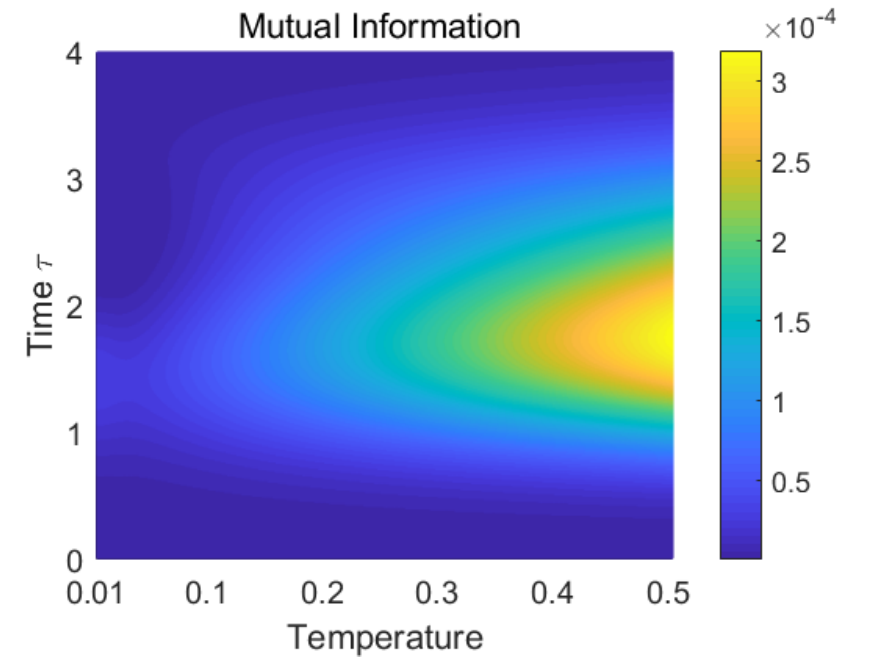}
}
\caption{(a) Variation of mutual information with acceleration and proper time. The acceleration of the detectors increases from $0.01$ while they accelerate with the same proper acceleration and are a distance $\pi$ apart. (b) Variation of mutual information with temperature of field and time. $T$ is scaled by multiplication by $2\pi$. The temperature of the detectors increases from $0.01$ while they remain stationary at a distance $\pi$ apart. All the parameters are the same as those in Figs.~\ref{fig:1a} and \ref{fig:1b}.}
\label{fig}
\end{figure}

The mutual information between the two detectors, which quantifies the total quantum correlation, is shown in Figs.~\ref{fig:2a} and \ref{fig:2b}. In our setting, the quantum mutual information has almost the same trends in time and acceleration (temperature) compared to the quantum discord. For simplicity, we consider only the quantum discord in detail, but the discussion holds also for the mutual information.

\begin{figure}[htbp!] \centering
\subfigure[ ] { \label{fig:3a}
\includegraphics[width=2.0in]{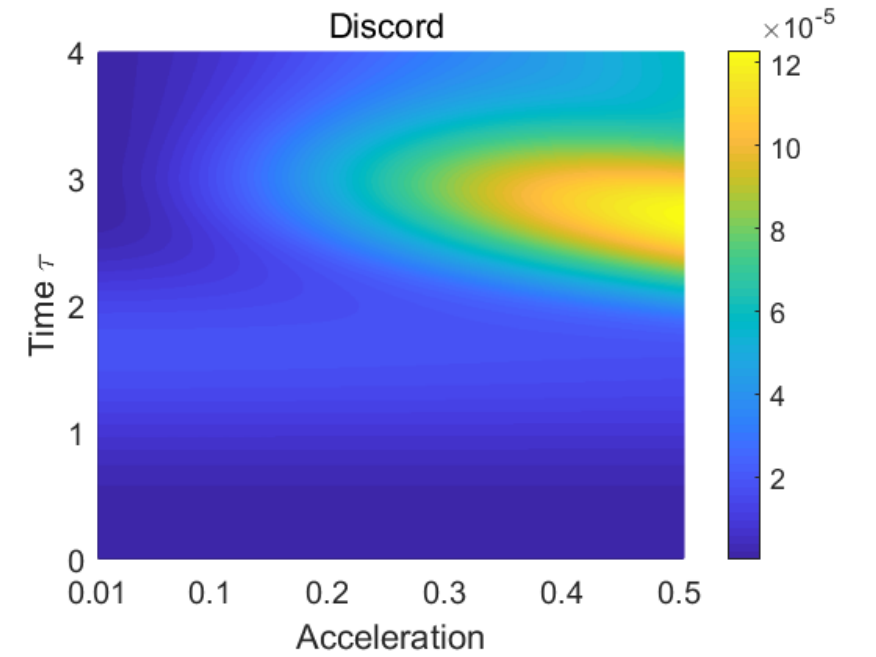}
}
\subfigure[ ] { \label{fig:3b}
\includegraphics[width=2.0in]{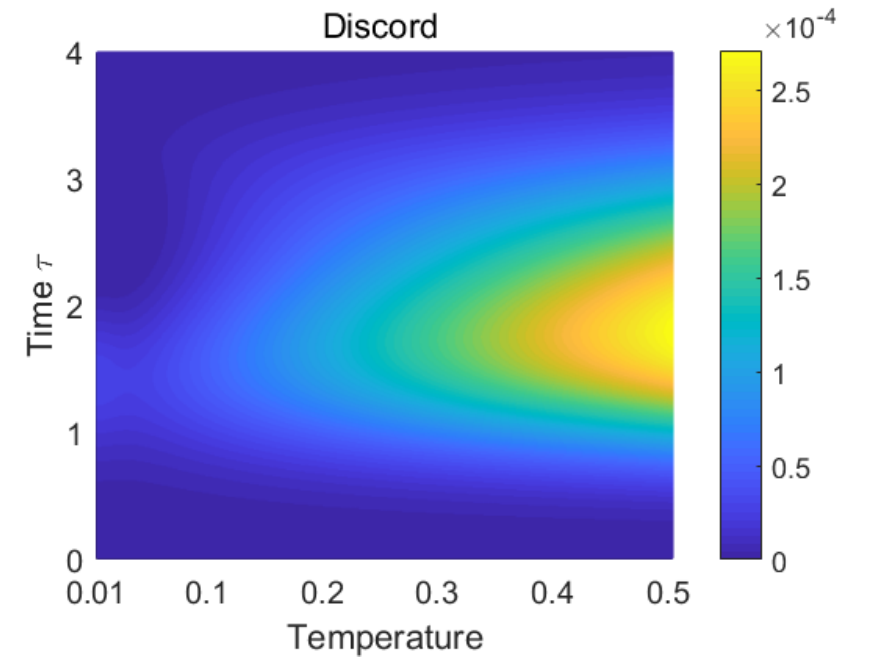}
}
\caption{(a) Variation of quantum discord with acceleration and proper time. The acceleration of the detectors increases from $0.01$ while the two detectors accelerate with the same proper acceleration and are a distance $\pi$ apart. (b) Variation of quantum discord with temperature and proper time. $T$ is scaled by multiplication by $2\pi$. The temperature of the detectors increases from $0.01$ while the two detectors remain stationary at a distance $\pi$ apart. All the parameters are the same as those in Figs.~\ref{fig:1a} and \ref{fig:1b}.}
\label{fig}
\end{figure}
In Figs.~\ref{fig:3a} and \ref{fig:3b}, we plot the quantum discord between the detectors in the same scenario as that in Figs.~\ref{fig:3a} and \ref{fig:3b}. Quantum discord is a measure of the pure quantum part of the correlations obtained by subtracting the classical correlations from the quantum mutual information. The trend in Figs.~\ref{fig:3a} and \ref{fig:3b} for the quantum discord is very similar to that for the quantum mutual information in Figs.~\ref{fig:2a} and \ref{fig:2b}. The quantum discord (mutual information) varies with the acceleration and time or the temperature and time. Their behaviors in time are quite similar to that of the entanglement $E_N$, and this follows the same explanation as that of the entanglement of $E_N$ with the quantum recurrence. Again we see that the quantum recurrence appears later in time for the ADs than for the stationary detectors in the thermal bath. The difference is that the quantum discord (mutual information) is amplified by the higher temperature rather than being reduced as in the case of entanglement.

At higher acceleration or temperature, the quantum discord (mutual information) is more robust than is the entanglement $E_N$. Moreover, as the acceleration (temperature) of the stationary detectors increases, the quantum discord can be amplified by increasing the temperature, which is consistent with the results in Ref.~\citenum{r23}. The entanglement and quantum discord have a monogamous relationship in a tripartite system (two subsystems in an environment)~\cite{r28}. Moreover, Ref.~\citenum{r29} shows that the evolution of the discord reflects the entanglement entropy dynamics between the system and the environment, and its behavior is opposite to that of the entanglement between the (sub)systems. This implies that increasing the temperature of the field can cause the detectors to be more mixed and therefore more likely to be entangled with the environment in our case. This leads to an increased amount of discord (mutual information) but less entanglement between the detectors. Note also that the discord (mutual information) enhancement is faster for the ADs than for the stationary detectors in a thermal bath.

Unlike the entanglement $E_N$, the quantum discord (mutual information) may exist for a long time and still fluctuate over time in Fig.~\ref{fig:2b}. We suggest that the evolution of the quantum discord (mutual information) in such systems can be understood as a competition between the decoherence effect among the systems due to the environment (field) and the system--environment (detector--field) entanglement that such decoherence also tends to generate. As we have seen, in this case the latter wins. The quantum discord is sustained because of the dominant system--environment (detector--field) entanglement over the decoherence or the disentanglement from the system (detector--detector). On the other hand, the system--environment entanglement does not boost the entanglement within or between the systems. Indeed, because of monogamy, the entanglement is usually further damaged and therefore more easily destroyed, as shown in Fig.~\ref{fig:1b}.

There are also apparent differences between the effects of the temperature of the flat space and the effects of the corresponding accelerations. Meanwhile, Ref.~\citenum{ref30} showed that the entanglement profiles of two accelerated qubit detectors and two stationary qubit detectors in a thermal field are largely the same but with certain differences. The correlation function plays an important role in quantifying the entanglement. Although a single AD is equivalent to a stationary detector in a thermal bath, this does not guarantee the equivalence of the quantitative correlation between two ADs and the correlation between two stationary detectors in a thermal bath because intuitively the latter detectors are surrounded by a global thermal bath while the former are surrounded locally by the associated local thermal baths~\cite{ref30,ref31}. Here we can see the similarities and differences between the effects of the temperature of the flat space and effects of the corresponding accelerations on the entanglement, quantum mutual information, and quantum discord. Unruh and Hawking pointed out that the equivalence of temperature and acceleration holds at single-observer level. Temperature and acceleration have certain different impacts on global correlations such as entanglement, mutual information, and discord. It is difficult for a detector in a sealed elevator to identify whether it is stationary in a thermal field or accelerating, but it is possible for two detectors to do so by examining their global correlation. For example, stationary detectors in a thermal field can access the quantum correlation faster than can ADs.

\subsection{Accelerations of detectors in opposite directions}

\begin{figure}[htbp!] \centering
\subfigure[ ] { \label{fig:4a}
\includegraphics[width=2.0in]{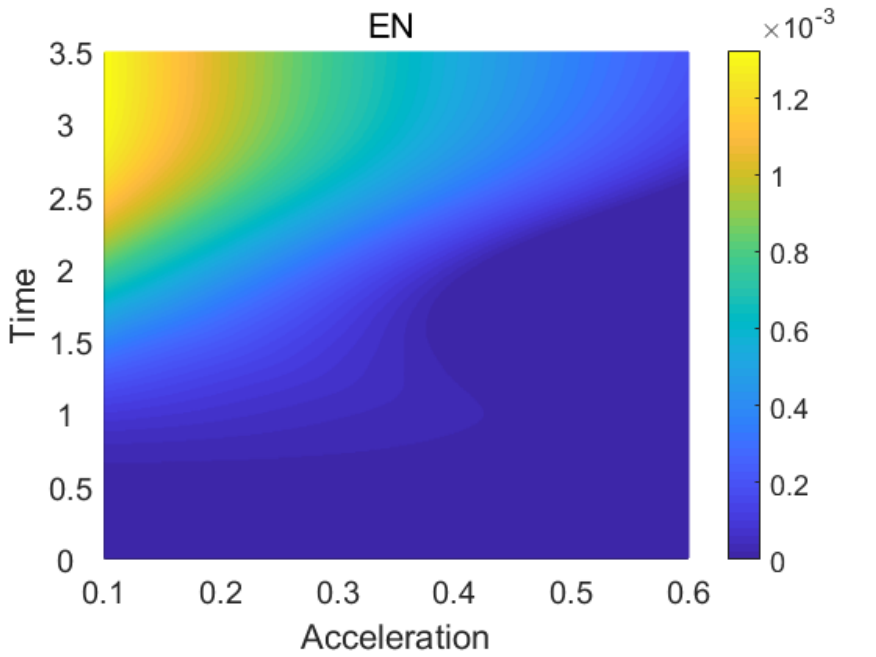}
}
\subfigure[ ] { \label{fig:4b}
\includegraphics[width=2.0in]{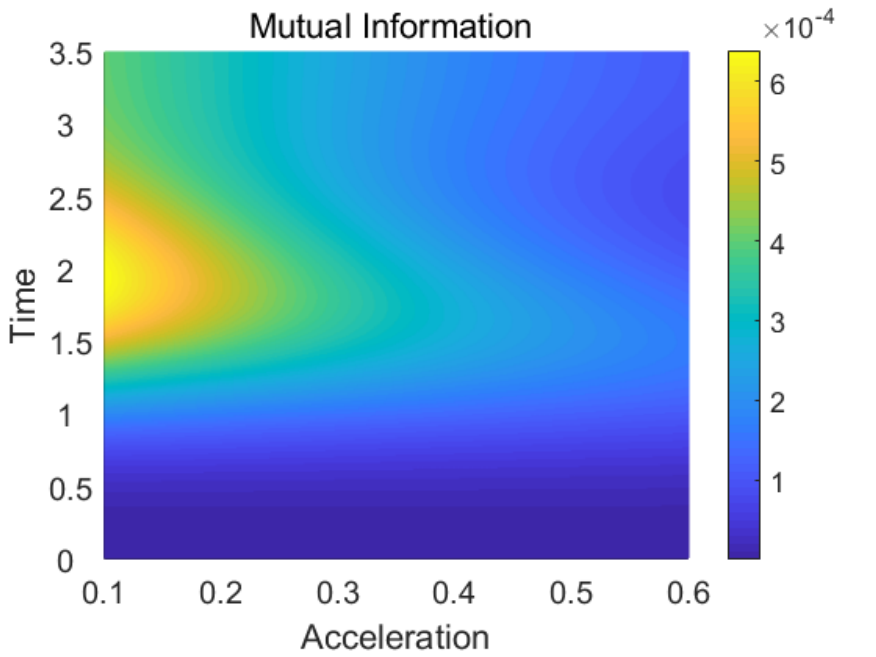}
}
\subfigure[ ] { \label{fig:4c}
\includegraphics[width=2.0in]{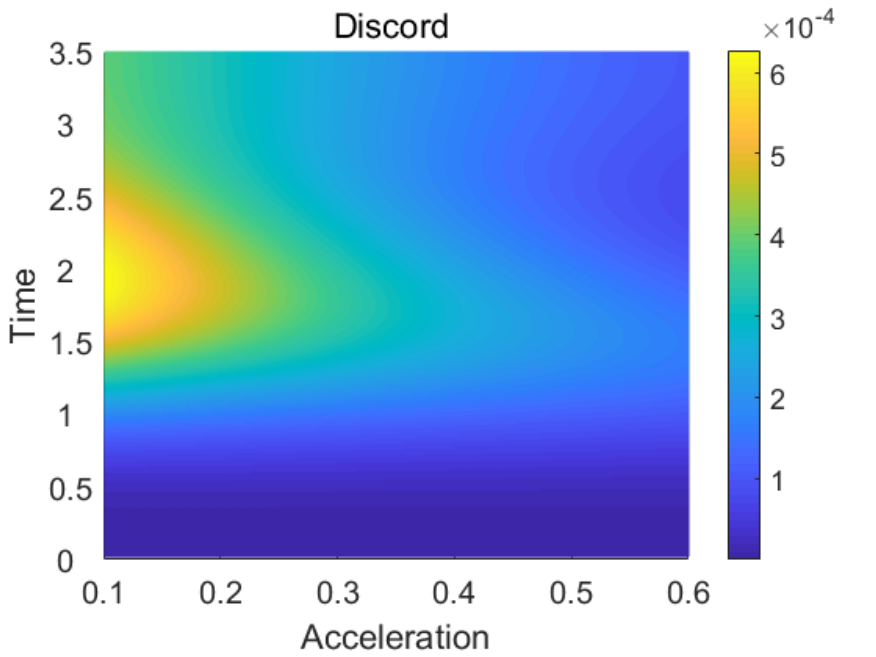}
}
\caption{Variation of (a) entanglement $E_{N}$, (b) quantum mutual information, and (c) quantum discord with acceleration and proper time $\tau$. The acceleration of the detectors increases from $0.1$ and the two detectors accelerate with the same proper acceleration in opposite directions starting at $2\pi$. The coupling strength for both detectors is set as $\lambda(\tau) =0.05\exp(-(\tau-1.5)^2)$. All the parameters are the same as those in Figs.~\ref{fig:1a} and \ref{fig:1b}.}
\end{figure}

We now investigate the entanglement, quantum mutual information, and quantum discord between two detectors accelerating in opposite directions. The world lines for Alice and Bob are $(a^{-1}\sinh(a\tau),(a^{-1}\cosh(a\tau)-1+2a\pi),0,0)$ and $(a^{-1}\sinh(a\tau),-a^{-1}(\cosh(a\tau)-1-2a\pi),0,0)$, respectively, which mimic two heat sources in flat spacetime moving away from each other. From Figs.~\ref{fig:4a}--\ref{fig:4c}, as time goes by, the two detectors are farther from each other at a constant acceleration; this is also true over acceleration at fixed time. They are all resurrected from dead over time, and the quantum correlation decreases monotonically over acceleration at fixed time. Once again, we see the Unruh effect of bipartition. The higher acceleration (temperature) causes the detectors to be more classical by lacking a quantum nature, thereby decreasing the entanglement. Because the larger acceleration means more-distant detectors at fixed time, the mutual information (discord) is not amplified by the higher acceleration but is instead reduced. This shows that the distance effect dominates the evolution. We suggest that the evolution of the quantum discord (mutual information) in such systems can be understood as a competition between the distance effect among the detectors due to the opposite accelerations and the thermal amplification that such accelerations tend to generate. As we have seen, in this case the former wins.

\section{\label{sec:Nonequilibrium} NONEQUILIBRIUM QUANTUM CORRELATIONS BETWEEN TWO DIFFERENTLY ACCELERATED DETECTORS}

\subsection{Nonequilibrium quantum correlations between equally accelerated detectors from coupling differences to scalar field}

\subsubsection{Quantum entanglement}

\begin{figure}[htbp!] \centering
\subfigure[ ] { \label{fig:5a}
\includegraphics[width=2.0in]{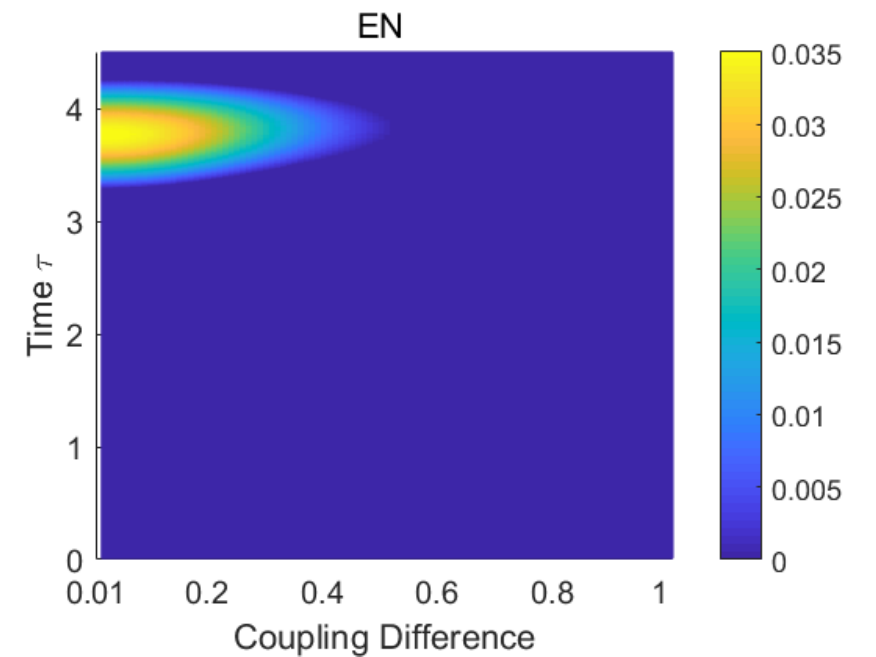}
}
\subfigure[ ] { \label{fig:5b}
\includegraphics[width=2.0in]{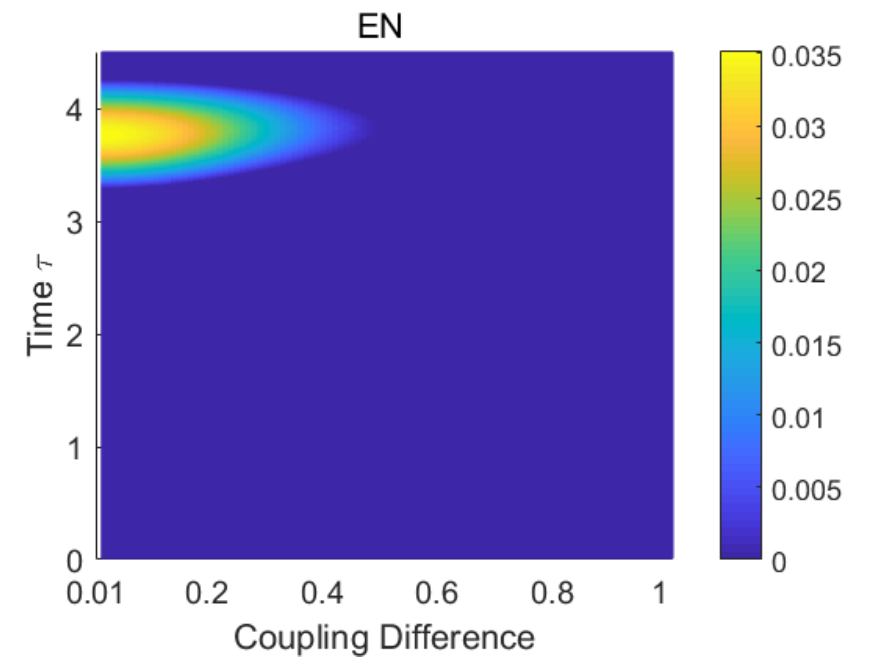}
}
\caption{(a) Variation of  $E_N$ with coupling difference and proper time. Alice and Bob remain at a lower constant proper acceleration of $0.01$ and a distance $\pi$ apart. (b) Variation of $E_N$ with coupling difference and time. Alex and Robb are stationary at thermal equilibrium at a lower temperature of $0.02\pi$ and also a distance $\pi$ apart. All the parameters are the same as those in Figs.~\ref{fig:1a} and \ref{fig:1b} except that $\lambda$ is constant with time.}
\end{figure}

There are two types of nonequilibrium in the system, one of which is caused by the imbalanced couplings of the two equally accelerated detectors (or detectors in a thermal field). If the couplings of the ADs to the scalar field are imbalanced, then the occupation or the number density for the detectors differs in time, which leads to an imbalance or nonequilibrium dynamics in time. However, when the system relaxes to a steady or stationary state at long times (if it can), one can show that this type of time-dependent nonequilibrium disappears~\cite{ref32,Ref32}. The other type of nonequilibrium is that if the two detectors are under different accelerations, then according to the Unruh effect, they are surrounded by environments with different temperatures. This type of nonequilibrium---which we refer to as the intrinsic nonequilibrium---exists even when the system relaxes to a steady state at long times. The temperature difference characterizes the degree of detailed balance-breaking and leads to the intrinsic nonequilibrium. In this sense, the latter is caused by the difference in accelerations or the difference in the corresponding temperatures. We first consider quantum correlations in different coupling strengths with equal accelerations because we seek to separate the pure nonequilibrium effects from that of the increased distance due to the accelerations. The world lines are given as $(t,\pi,0,0)$ and $(t,2\pi,0,0)$ for the two stationary detectors Alex and Robb, respectively, and as $(a^{-1}\sinh(a\tau),a^{-1} (\cosh(a\tau)-1),0,0)$ and $(a^{-1}\sinh(a\tau),a^{-1}(\cosh(a\tau)-1+a\pi),0,0)$ for the two ADs Alice and Bob, respectively. The coupling is assumed to be constant over time, and we set $\lambda_{1}+\lambda_{2}$ to a constant. The entanglement recurs and dies again over time, similar to the equilibrium case. At fixed proper times, we can see this nontrivial nonequilibrium effect through the coupling imbalance, which decreases the entanglement between the ADs in Fig.~\ref{fig:5a} and that between the stationary detectors in a thermal bath in Fig.~\ref{fig:5b}. Their behaviors are very similar, and this nontrivial dependence of entanglement versus nonequilibrium may be due to the competition between coherence and population redistribution~\cite{ref32}.

\subsubsection{Quantum mutual information and quantum discord}

\begin{figure}[htbp!] \centering
\subfigure[ ] { \label{fig:6a}
\includegraphics[width=2.0in]{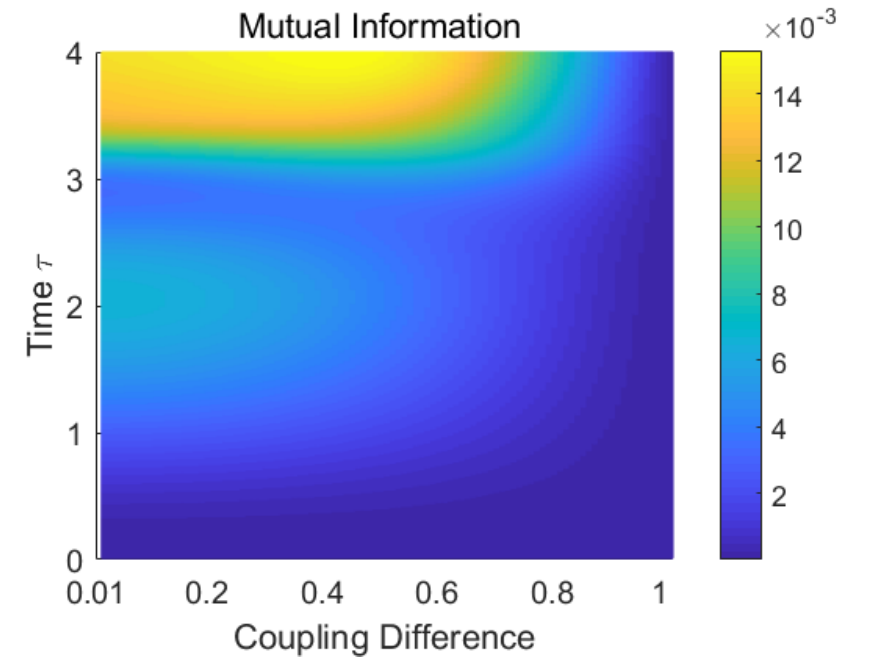}
}
\subfigure[ ] { \label{fig:6b}
\includegraphics[width=2.0in]{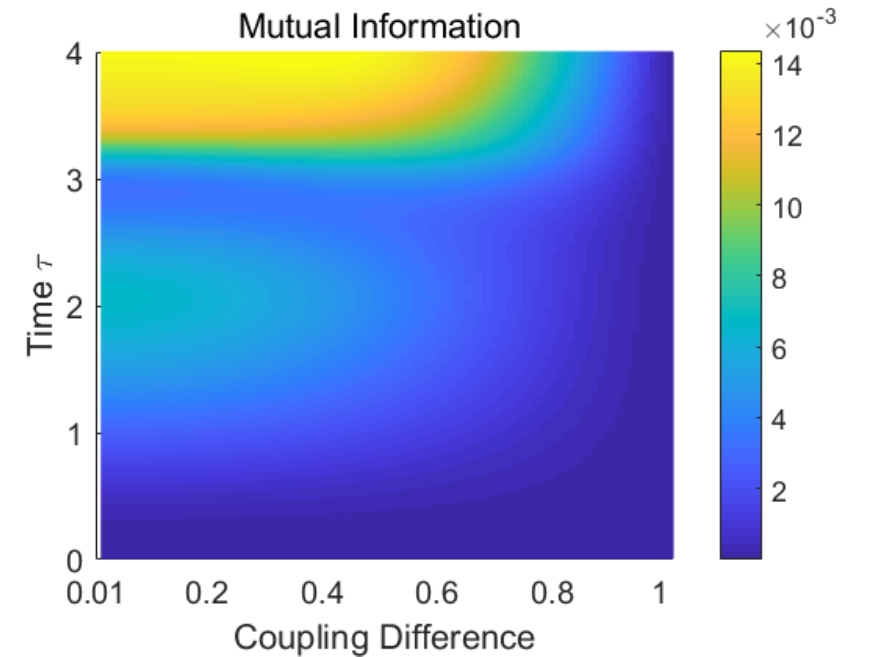}
}
\caption{(a) Variation of quantum mutual information with coupling difference and proper time. Alice and Bob remain at a lower constant proper acceleration of $0.01$ and a distance $\pi$ apart. (b) Variation of mutual information with coupling difference and time. Alex and Robb are stationary at thermal equilibrium at a lower temperature of $0.02\pi$ and also a distance $\pi$ apart. All the parameters are the same as those in Figs.~\ref{fig:1a} and \ref{fig:1b}.}
\end{figure}

\begin{figure}[htbp!] \centering
\subfigure[ ] { \label{fig:7a}
\includegraphics[width=2.0in]{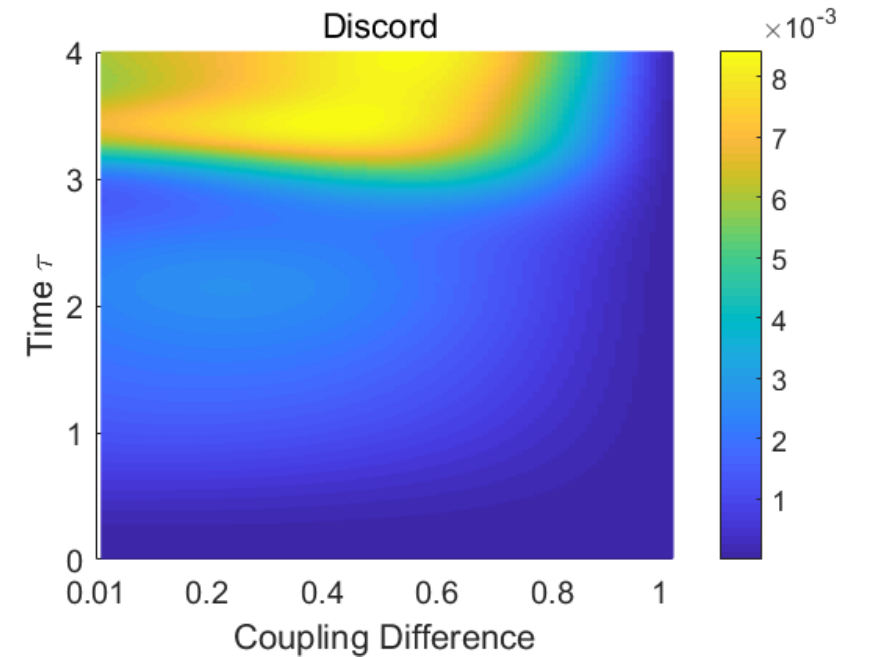}
}
\subfigure[ ] { \label{fig:7b}
\includegraphics[width=2.0in]{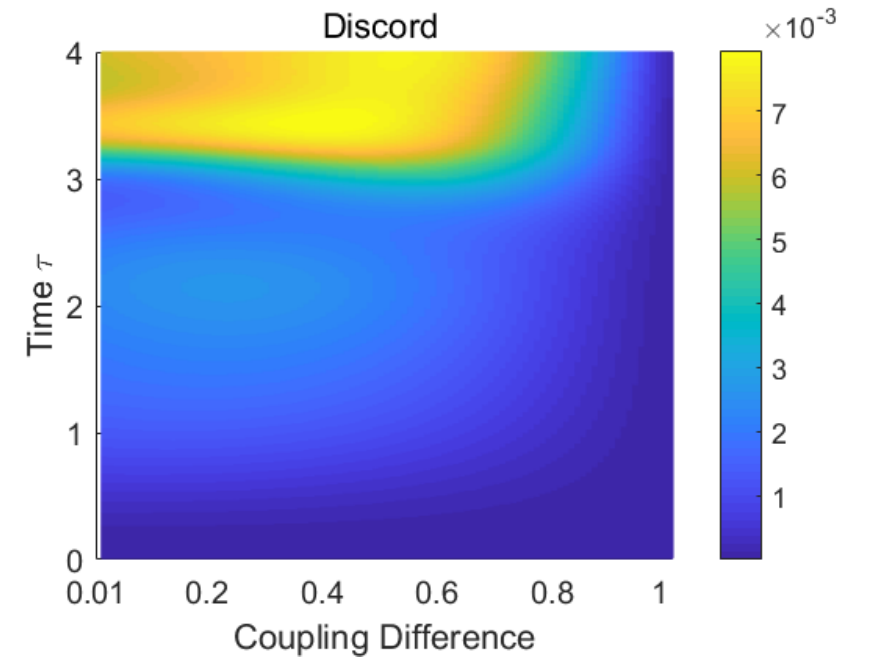}
}
\caption{(a) Variation of quantum discord with coupling difference and proper time while Alice and Bob remain at a lower constant proper acceleration of $0.01$ and a distance $\pi$ apart. (b) Variation of quantum discord with coupling difference and time while Alex and Robb are stationary at thermal equilibrium at a lower temperature of $0.02\pi$ and also a distance $\pi$ apart. All the parameters are the same as those in Figs.~\ref{fig:1a} and \ref{fig:1b}.}
\end{figure}

Figures~\ref{fig:6a}, \ref{fig:6b}, \ref{fig:7a}, and \ref{fig:7b} show the quantum mutual information and discord with respect to the coupling difference and time at the same temperature and the same acceleration. Their behaviors are very similar, and in our setting we find that the quantum mutual information has almost the same trend as that of the quantum discord. For simplicity, we discuss only the quantum discord in detail, but the discussion holds also for the quantum mutual information.

The quantum mutual information and quantum discord behave nonmonotonically in time and exhibit recurrent behavior. At fixed proper time, the quantum mutual information and quantum discord can be enhanced by the coupling difference in a large coupling-difference range. Eventually, the quantum mutual information and quantum discord can be reduced as the coupling difference becomes large. These behaviors lead to a wide coupling-difference zone for the detectors to harvest the quantum correlation from the nonequilibrium scenario. Similar behaviors were also observed in Ref.~\citenum{ref32} for the quantum correlations between two qubits (stationary detectors) in nonequilibrium environments as the result of the competition between coherence and population.
The enhancement of the quantum correlations is apparent from the higher coupling difference or nonequilibrium nature under the higher local accelerations, and here we rationalize that enhancement through the nonequilibrium nature. The latter creates a net energy flow to the system, much like a voltage generates an electric current. Because the flow is global in state space, the enhanced flow from the nonequilibrium nature can increase the global correlations~\cite{Ref32,Ref33}.

\subsection{Nonequilibrium quantum correlations between detectors with different accelerations}

In this part, we consider the case in which the two detectors in the cavity are in a nonequilibrium state with different accelerations. The world lines are $(a^{-1}\sinh(a\tau),a^{-1}(\cosh(a\tau)-1),0,0)$ and $(b^{-1}\sinh(b\tau),b^{-1}(\cosh(b\tau)-1),0,0)$ for the two detectors Alice and Bob, respectively.

Because there are different proper times associated with the accelerations, we choose the perspective of the detector with the larger acceleration to weaken the redshift effect. Meanwhile, in Appendix~\ref{sec:Appendix:C}, we provide plots that show the significant redshift effect in coordinate time. The quantum correlations are resurrected over time and are amplified by the acceleration difference, and the latter can be viewed as the measure of the nonequilibrium nature (the temperature difference through the Unruh effect). The acceleration difference also changes the distance between the two detectors, therefore they are prone to two effects, i.e., the larger distance and the larger acceleration difference, both of which influence the quantum correlation. The nonequilibrium nature amplifies the quantum correlations, while the larger distance weakens them. The amplification of the quantum correlations can be viewed as a result of the competition between the nonequilibrium nature and the distance, and we see that the former dominates the latter regarding the quantum correlations.

\begin{figure}\centering
\subfigure[ ] { \label{fig:8a}
\includegraphics[width=2.0in]{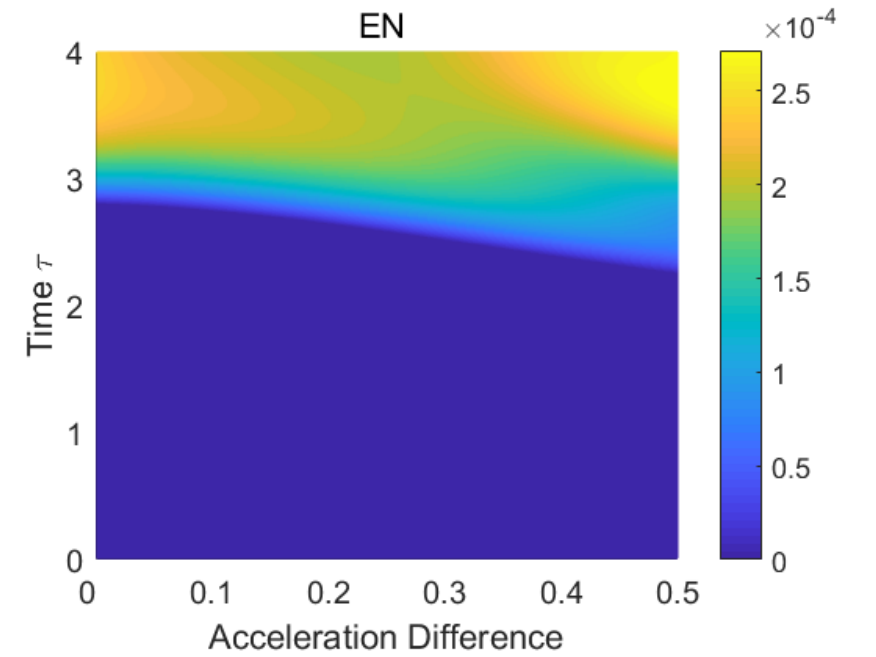}
}
\subfigure[ ] { \label{fig:8b}
\includegraphics[width=2.0in]{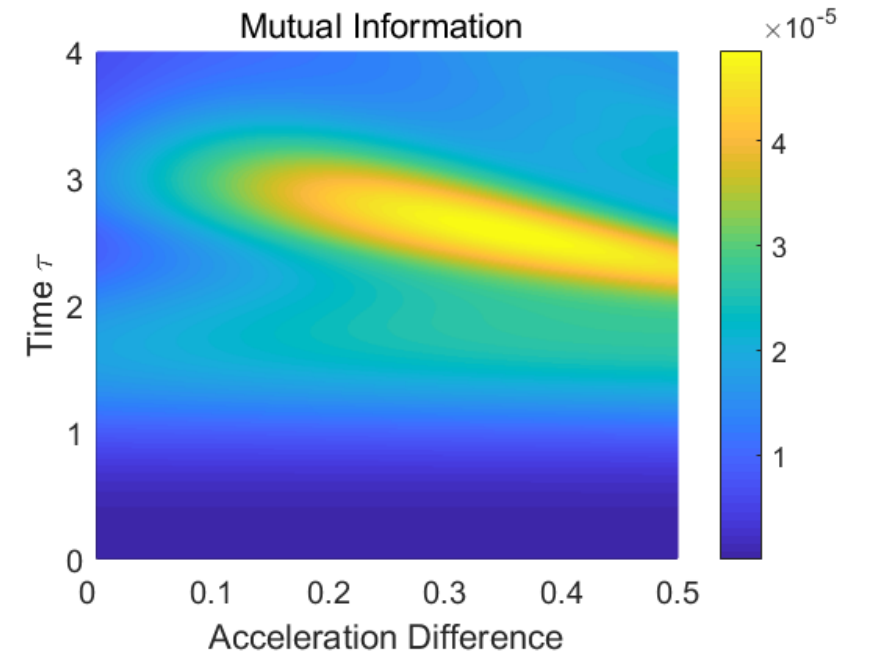}
}
\subfigure[ ] { \label{fig:8c}
\includegraphics[width=2.0in]{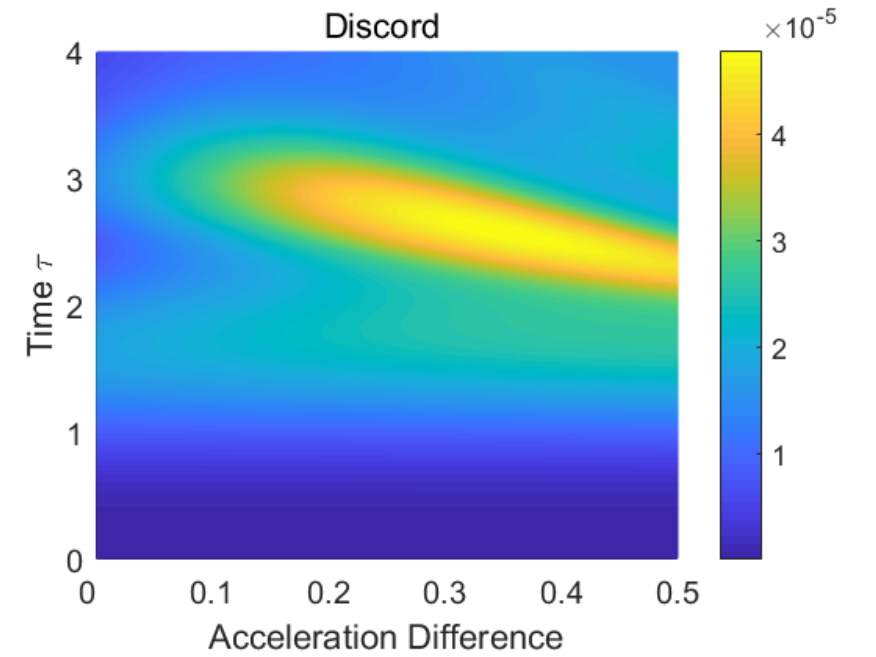}
}
\caption{Variation of (a) $E_N$, (b) mutual information, and (c) quantum discord with acceleration difference and Bob's proper time $\tau$. Alice remains at a lower acceleration $0.1$ and Bob has a larger acceleration starting from $0.1$. All the parameters are the same as those in Figs.~\ref{fig:1a} and \ref{fig:1b}.}
\end{figure}

\section{\label{sec:level5} CONCLUSION}

Using the developed methods, we discussed how equilibrium quantum correlations such as entanglement, mutual information, and quantum discord differ between (i) two equally accelerated detectors and (ii) two stationary detectors in a thermal field with temperature $T$. First, we studied quantum correlation harvesting from the field over time in several different scenarios~\cite{r18,r19,r22,r23}. The information flowing back from the system to the environment may lead to nonMarkovian dynamics~\cite{ref27,ref28,Ref27,Ref29}. The quantum correlation of bipartition has qualitatively similar trends regarding acceleration and temperature, as shown by the similar effects from the two detectors under the same acceleration and temperature. As the acceleration (temperature) of the stationary detectors increases, the entanglement disappears while the mutual information and the quantum discord increase. The reason for this was given as follows: the quantum discord (mutual information) is sustained because of the dominant system--environment (detector--field) entanglement over the decoherence or the disentanglement from the system (detector--detector). On the other hand, the system--environment entanglement does not boost the entanglement within or between the systems. Indeed, because of monogamy, the entanglement is usually further damaged. We also observed that the recurrence of quantum correlations appears later in time for ADs than for thermal stationary detectors. At fixed time, the entanglement reduction is faster for stationary detectors in a thermal bath than for ADs. Also studied was the quantum correlation between detectors accelerated in opposite directions. Remarkably, the quantum correlations decrease monotonically with the acceleration. We suggested that the evolution of quantum discord (mutual information) in such systems can be understood as a competition between (i) the distance between the detectors due to their opposite accelerations and (ii) the thermal amplification that such acceleration also tends to generate. As we have seen, in this case the former wins.

Meanwhile, we also studied nonequilibrium quantum correlations. We first considered the nonequilibrium effect in time caused by different couplings. We noted that the nonequilibrium nature in time influenced by the imbalanced couplings can strengthen the quantum correlations. This leads to a wide coupling-difference zone for harvesting the quantum correlations for nonequilibrium scenarios. Another nonequilibrium case that we considered was that of two ADs with different accelerations. This is the case of intrinsic nonequilibrium because the detailed balance is broken when the system relaxes to a steady state at long times. Besides, the acceleration difference leads to both the nonequilibrium effect (different equivalent temperatures) and the distance effect (farther from each other) on the quantum correlations between the two detectors. The nonequilibrium effect appears to dominate the distance effect in terms of enhancing the quantum correlations. To rationalize the nonequilibrium-enhanced quantum correlations, we noted that the nonequilibrium nature creates a net energy flow to the system, much like a voltage generates an electric current. Because the flow is global in state space, this helps the connections between states in distance, thereby decreasing the effective distances between states for communication. Therefore, we expect the nonequilibrium nature to enhance the quantum correlations through the global energy flow, or in other words effectively reducing the distances between states.

\begin{acknowledgments}
He Wang acknowledges the support of the National Natural Science Foundation of China (Grant No.\ 21721003) and the Ministry of Science and Technology of China (Grant No.\ 2016YFA0203200).
\end{acknowledgments}

\appendix

\section{\label{sec:Appendix:A} Derivation of $F(t)$}

For a general time-dependent quadratic Hamiltonian, it is convenient to express it in the form of an annihilation (creation) operator vector, which is defined as
\begin{equation}
\label{eq:31}
\hat{a}=[\hat{a}_{d_{1}},\hat{a}_{d_{2}},...\hat{a}_{d_{j}}...\hat{a}_{1},\hat{a}_{2},...\hat{a}_{n}]^{T},
\end{equation}
\begin{equation}
\label{eq:32}
\hat{a}^{\dagger}=[\hat{a}_{d_{1}}^{\dagger},\hat{a}_{d_{2}}^{\dagger},...\hat{a}_{d_{j}}^{\dagger}...\hat{a}_{1}^{\dagger},\hat{a}_{2}^{\dagger},...\hat{a}_{n}^{\dagger}]^{T}.
\end{equation}
A general quadratic Hamiltonian such as Eq.~\eqref{eq:11} can be written as
\begin{equation}
\label{eq:33}
\hat{H}^{H}=(\hat{a}^{\dagger})^{T}w(t)\hat{a}+(\hat{a}^{\dagger})^{T}m(t)\hat{a}^{\dagger}+\hat{a}^{T}m(t)^{H}\hat{a},
\end{equation}
where $w(t)$ and $m(t)$ are the coefficient matrices, and $m(t)^{H}$ is the conjugate transpose of $m(t)$, i.e., $m(t)^{H}=m(t)^{T*}$. Combining Eqs.~\eqref{eq:15} and \eqref{eq:33} gives the matrix form of $F(t)$ as
\begin{equation}
\label{eq:34}
F(t)=\begin{bmatrix} A(t) & C{t} \\ C(t)^{H} & B(t) \end{bmatrix},
\end{equation}
where
\begin{equation}
\label{eq:35}
A(t)=\frac{1}{2}(w(t)+m(t)+m(t)^{H}),
\end{equation}
\begin{equation}
\label{eq:36}
B(t)=\frac{1}{2}(w(t)-m(t)-m(t)^{H}),
\end{equation}
\begin{equation}
\label{eq:37}
C(t)=\frac{1}{2}(w(t)-m(t)+m(t)^{H}).
\end{equation}

\section{\label{sec:Appendix:B} Causality analysis of UV cutoff}

We aim to study the convergence of the finite modes used in the computation. Because the detectors are in a stationary cavity, the maximum Lorentz factor for a uniformly accelerated detector is $\gamma_{max}=1+aL$. Therefore, the detector does not sweep across many field modes because it is blue/red-shifted when $\Omega\to\gamma_{max}\Omega$. In our setup, we use a coupling with a Gaussian time profile, which requires far fewer modes for convergence and produces negligible switching noise compared to sharp-switching coupling~\cite{r22}. (Rather, a sharp $\lambda(\tau)$ can excite field modes significantly higher than those near resonance. This is because the off-resonant rotating-wave terms become important (as well as the counter-rotating wave terms) if the interaction changes suddenly in a characteristic time of less than $\sim$$1/\Omega$~\cite{Ref34}.) The Gaussian coupling also causes the ADs to be decoupled gradually from the field such that they only sweep across some low-frequency field modes. Taken together, these aspects suggest that not too many cavity modes are needed for convergence. In our setup, we use $80$ modes, a number that was chosen so that increasing $N$ further does not alter the results perceivably. In what follows, we show the convergence of the finite modes.

The cavity length naturally limits the lowest energy of the mode with the longest wavelength. A single-mode approximation in the Unruh--DeWitt model can lead to unphysical results such as broken causality~\cite{ref33}. For convergent results, one should consider infinitely many modes, but that is impossible for practical calculations, so instead a UV cutoff is necessary. How can a reasonable UV cutoff be chosen? In Ref.~\citenum{ref27}, an appropriate cutoff was selected based on whether the accelerator strictly produces the corresponding Unruh temperature, and a larger cutoff corresponds to an exact Unruh effect $T=a/2\pi$. In the present study, we focus on quantifying the quantum correlations between bipartite ADs and bipartite stationary detectors in a thermal field. Some studies have indicated that the effect of the acceleration is not always equivalent to that of the temperature on the bipartite effect (such as the Casimir force)~\cite{ref35}. It is difficult to find the appropriate regime for checking the equivalency between the acceleration effect and the temperature effect based on only the behaviors of the individual detectors. This is because even when a single detector does not show different behaviors from acceleration and temperature, bipartite correlations may reflect the difference. Therefore, we should consider the results for the bipartite correlations in addition to those from a single detector to establish the differences in the effects. In this situation, we consider another criterion that is based on the degree of causality-breaking: a real effective physical model must be causal for any two separated parts. It has been shown that a finite number of modes can lead to superluminal signaling, but nevertheless it is possible to have a finite number of modes without acausal signaling appearing within the required arbitrarily high precision for convergence~\cite{r22}.

\begin{figure}[htbp]
\subfigure[ ] { \label{fig:9a}
\includegraphics[width=2.0in]{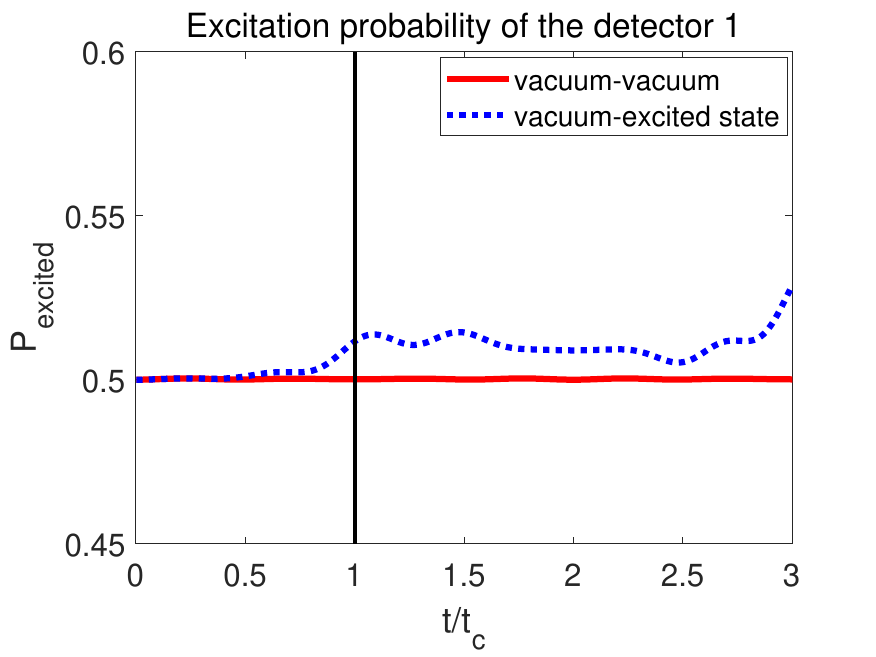}
}
\subfigure[ ] { \label{fig:9b}
\includegraphics[width=2.0in]{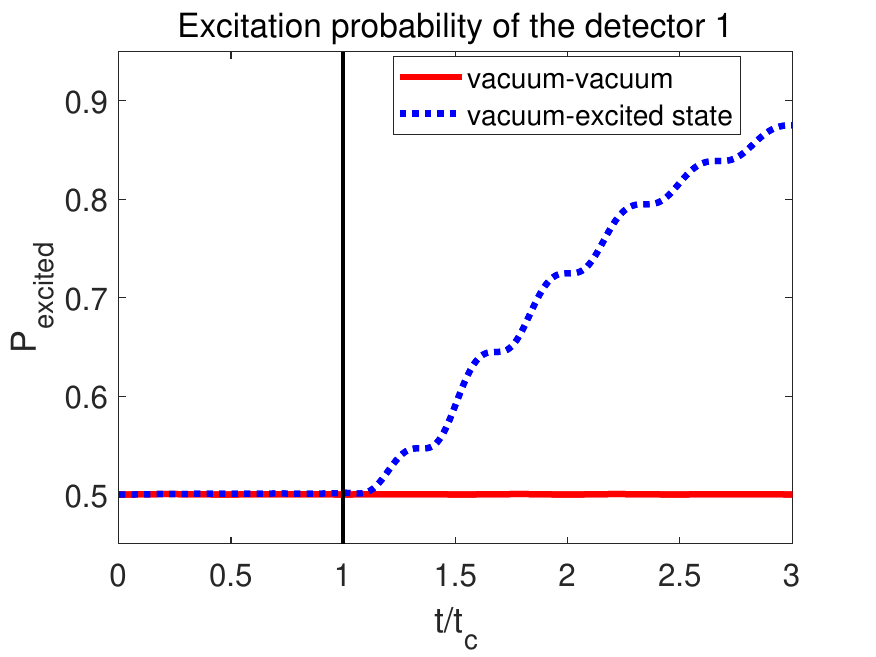}
}
\caption{Excitation probability as a function of $t/t_{c}$ of an inertial detector (detector~1) in the presence of another in a highly excited state (detector~2) considering (a) 7 and (b) 10 modes. The vertical line represents the time $t=t_{c}$ at which the two detectors come into causal contact. In each plot, the blue dashed curve corresponds to the excitation probability for detector~1 when detector~2 is initially in its vacuum state, whereas for the solid (red) curve we initialize detector~2 in a squeezed state with squeezing parameter $r=5$. In (a), the blue dashed line deviates from the red solid line before the two detectors come into causal contact, whereas in (b) the blue and red lines conform before the two detectors come into causal contact. Notably, there are also certain slight deviations in (b). The parameter values are $L=4\pi$, $\lambda = 0.05$, and $\Omega=3/2$. Detector~1 is at $\pi$ and detector~2 is at $3\pi$. Periodic boundary conditions are used here.}
\end{figure}

Consider the following setup: detectors~1 and 2 are stationary and separated by a distance in a cavity. Detector~1 is in the vacuum state, while detector~2 is in either the vacuum or excited state. We want to examine how long it takes detector~1 to observe the effects of detector~2 via propagation of the excitation in the field. The excitation probability of detector~1 is given as~\cite{ref34}
\begin{equation}
\label{eq:38}
p=1-\frac{2}{\sqrt{det(\sigma_{d_{1}})+Tr(\sigma_{d_{1}})+1}}.
\end{equation}
Furthermore, we take the highly excited state of detector~2 to be specifically a single-mode squeezed state with a covariance matrix of the form
\begin{equation}
\label{eq:39}
\sigma_{d_{2}}=\begin{bmatrix} e^{2r} & 0 \\ 0 & e^{-2r} \end{bmatrix},
\end{equation}
where $r$ is the squeezing parameter. In Figs.~\ref{fig:9a} and \ref{fig:9b}, we plot the excitation probability of detector~1 as a function of $t/t_{c}$, where $t_{c}$ is the time at which the two detectors come into causal contact. We show the results for $7$ and $10$ field modes, and the vertical lines in Fig.~9 represent the time $t=t_{c}$. In each plot, the blue dashed curve corresponds to the excitation probability for detector~1 when detector~2 is initially in its vacuum state. For the solid (red) curve, we initialize detector~2 in a squeezed state with squeezing parameter $r=5$. Increased excitations in detector~1 caused by the propagating field quanta emitted from squeezed detector~2 are observed as expected. We also observe that if not enough field modes are used, then the additional excitation occurs before the two detectors are in causal contact, as in Fig.~\ref{fig:9a}. It is only when enough field modes are included that the two curves start to diverge at $t=t_{c}$, as in Fig.~\ref{fig:9b}. This shows when they start to have mutual influence. Notably, there are also certain slight deviations in Fig.~\ref{fig:9b}. Upon increasing the number of modes used, the deviation before causal contact becomes extremely small until the specified precision. Considering that physical UV and IR cutoffs must guarantee causality, we examine the causality to determine whether a mode number is appropriate. 

\section{\label{sec:Appendix:C} Nonequilibrium quantum correlations in coordinate time}

\begin{figure}[h] \centering
\subfigure[] { \label{fig:11a}
\includegraphics[width=1.7in]{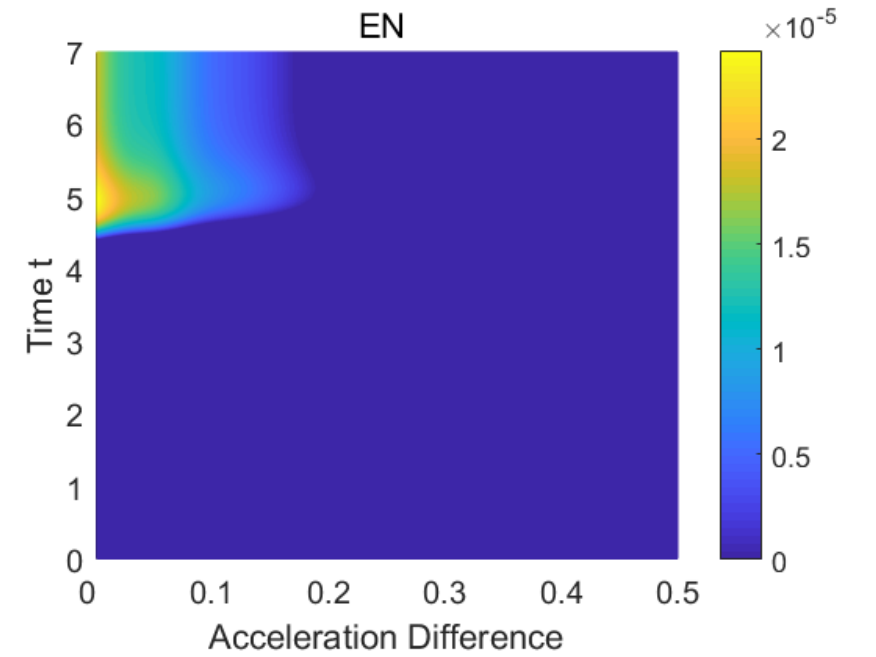}
}\subfigure[] { \label{fig:11b}
\includegraphics[width=1.7in]{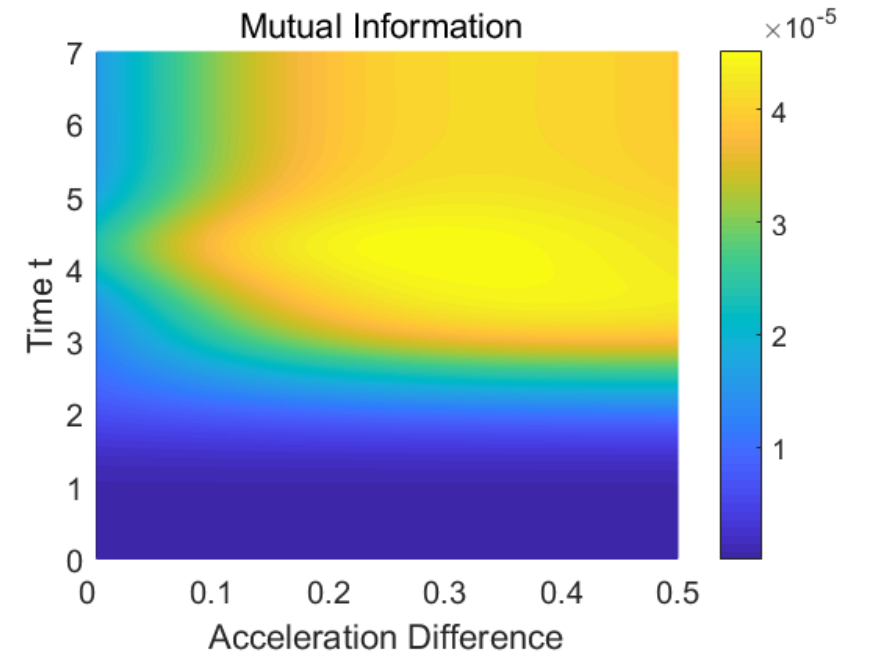}
}
\subfigure[] { \label{fig:11c}
\includegraphics[width=1.7in]{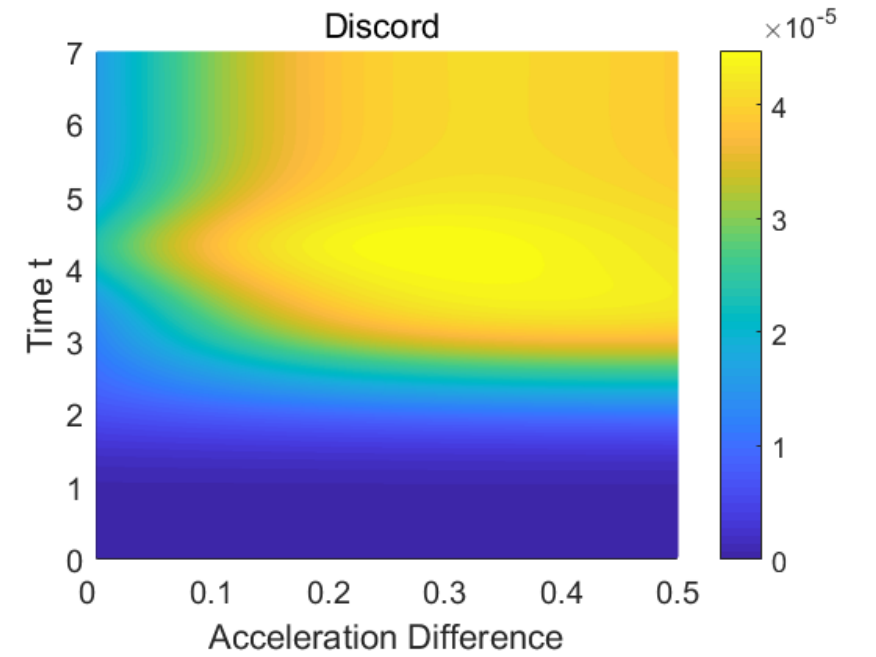}
}
\caption{Variation of (a) $E_{N}$, (b) mutual information, and (c) quantum discord with acceleration difference and coordinate time $t$. Alice remains at a lower acceleration of $0.1$ and Bob has a larger acceleration starting from $0.1$. All the parameters are the same as those in Figs.~1(a) and 1(b). Here we see a significant redshift in coordinate time, and the period is increased.}
\end{figure}

Figures~\ref{fig:11a}--\ref{fig:11c} show the quantum correlations with respect to coordinate time and acceleration difference. The entanglement decreases but the mutual information (discord) is amplified by the higher acceleration difference. Here we see a significant redshift in coordinate time, and the period is increased.

\clearpage
\nocite{*}


\begin{thebibliography}{99}

\bibitem{r1}
M.\ A.\ Nielsen and I.\ L.\ Chuang, \textit{Quantum Computation and Quantum Information: 10th Anniversary Edition} (Cambridge University Press, Cambridge, UK, 2011).

\bibitem{r2}
A.\ Einstein, B.\ Podolsky, and N.\ Rosen, Can quantum-mechanical description of physical reality be considered complete?, Phys.\ Rev.\ {\bf 47}, 777 (1935).

\bibitem{r3}
R.\ Horodecki, P.\ Horodecki, M.\ Horodecki, \textit{et~al.}, Quantum entanglement, Rev.\ Mod.\ Phys.\ {\bf 81}, 865 (2009).

\bibitem{r4}
V.\ P.\ Belavkin and M.\ Ohya, Entanglement, quantum entropy and mutual information, Proc.\ R.\ Soc.\ A {\bf 458}, 209 (2002).

\bibitem{r5}
H.\ Ollivier and W.\ H.\ Zurek, Quantum discord: A measure of the quantumness of correlations, Phys.\ Rev.\ Lett.\ {\bf 88}, 017901 (2001).

\bibitem{r6}
G.\ Adesso, I.\ Fuentes-Schuller, and M.\ Ericsson, {Continuous variable entanglement sharing in non-inertial frames}, Phys.\ Rev.\ A {\bf 76}(6), 62112 (2007).

\bibitem{r7}
I.\ Fuentes-Schuller and R.\ B.\ Mann, {Alice falls into a black hole: Entanglement in non-inertial frames}, Phys.\ Rev.\ Lett.\ {\bf 95}(12), 120404 (2004).

\bibitem{r8}
P.\ M.\ Alsing, I.\ Fuentes-Schuller, R.\ B.\ Mann, \textit{et~al.}, {Entanglement of Dirac fields in non-inertial frames}, Phys.\ Rev.\ A {\bf 74}(3), 396 (2006).

\bibitem{r9}
J.\ L.\ Ball, I.\ Fuentes-Schuller, and F.\ P.\ Schuller, {Entanglement in an expanding spacetime}, Phys.\ Lett.\ A {\bf 359}(6), 550 (2006).

\bibitem{r10}
Y.\ Ling, S.\ He, W.\ Qiu, \textit{et~al.}, {Quantum entanglement of electromagnetic field in non-inertial reference frames}, J.\ Phys.\ A {\bf 65}(30), 9025 (2007).

\bibitem{r11}
E.\ Martin-Martinez and N.\ C.\ Menicucci, {Cosmological quantum entanglement}, Class.\ Quantum Grav.\ {\bf 29}(22), 224003 (2012).

\bibitem{r12}
E.\ Martin-Martinez and N.\ C.\ Menicucci, {Entanglement in curved spacetimes and cosmology}, Class.\ Quantum Grav.\ {\bf 31}(21), 214001 (2014).

\bibitem{r13}
W.\ G.\ Unruh, {Notes on black hole evaporation}, Phys.\ Rev.\ D {\bf 14}(4), 870 (1976).

\bibitem{r14}
S.\ W.\ Hawking, {Particle creation by black holes}, Commun.\ Math.\ Phys.\ {\bf 43}, 199 (1975).

\bibitem{r15}
S.\ S.\ Schweber and J.\ C.\ Polkinghorne, {An introduction to relativistic quantum field theory}, Phys.\ Today {\bf 15}(3), 66 (1962).

\bibitem{ref16}
D.\ E.\ Bruschi, J.\ Louko, E.\ Martin-Martinez, \textit{et~al.}, {The Unruh effect in quantum information beyond the single-mode approximation}, Phys.\ Rev.\ A {\bf 82}, 042332 (2010).

\bibitem{r16}
B.\ DeWitt, \textit{General Relativity; an Einstein Centenary Survey} (Cambridge University Press, Cambridge, UK, 1980).

\bibitem{r17}
L.\ C.\ B.\ Crispino, A.\ Higuchi, and G.\ E.\ A.\ Matsas, {The Unruh effect and its applications}, Rev.\ Mod.\ Phys.\ {\bf 80}(3), 787 (1975).

\bibitem{r18}
E.\ Martin-Martinez, E.\ G.\ Brown, W.\ Donnelly, \textit{et~al.}, {Sustainable entanglement production from a quantum field}, Phys.\ Rev.\ A {\bf 88}(5), 11592 (2013).

\bibitem{r19}
A.\ Sachs, R.\ B.\ Mann, and E.\ Martin-Martinez, {Entanglement harvesting and divergences in quadratic Unruh--DeWitt detector pairs}, Phys.\ Rev.\ D {\bf 96}(8), 080512 (2017).

\bibitem{r20}
S.\ Y.\ Lin, C.\ H.\ Chou, and B.\ L.\ Hu, {Disentanglement of two harmonic oscillators in relativistic motion}, Phys.\ Rev.\ D {\bf 78}(12), 667 (2008).

\bibitem{r21}
S.\ Y.\ Lin and B.\ L.\ Hu, {Entanglement creation between two causally-disconnected objects}, Phys.\ Rev.\ D {\bf 81}(4), 389 (2009).

\bibitem{r22}
E.\ G.\ Brown, E.\ Martin-Martinez, N.\ C.\ Menicucci, \textit{et~al.}, {Detectors for probing relativistic quantum physics beyond perturbation theory}, Phys.\ Rev.\ D {\bf 87}(8), 084062 (2013).

\bibitem{r23}
E.\ G.\ Brown, Thermal amplification of field correlation harvesting, Phys.\ Rev.\ A {\bf 88}(6), 062336 (2013).

\bibitem{r24}
G.\ Adesso, S.\ Ragy, and A.\ R.\ Lee, {Continuous variable quantum information: Gaussian states and beyond}, Open Syst.\ Inf.\ Dyn.\ {\bf 21}(01n02), 1440001 (2014).

\bibitem{r25}
G.\ Adesso, F.\ Illuminati, W.\ Donnelly, \textit{et~al.}, {Entanglement in continuous variable systems: Recent advances and current perspectives}, J.\ Phys.\ A {\bf 40}(28), 7821 (2007).

\bibitem{r26}
G.\ Adesso and A.\ Datta, {Quantum versus classical correlations in Gaussian states}, Phys.\ Rev.\ Lett.\ {\bf 105}(3), 030501 (2010).

\bibitem{ref27}
S.\ Vriend, D.\ Grimmer, and E.\ Martin-Martinez, {The Unruh effect in slow motion}, arxiv:2011.08223 (2020).

\bibitem{Universality28}
W.\ G.\ Brenna, E.\ G.\ Brown, R.\ B.\ Mann, and E.\ Martin-Martinez, {Universality and thermalization in the Unruh effect}, Phys.\ Rev.\ D {\bf 88}, 064031 (2013).

\bibitem{r27}
B.\ Bellomo, R.\ L.\ Franco, and G.\ Compagno, {Non-Markovian effects on the dynamics of entanglement}, Phys.\ Rev.\ Lett.\ {\bf 99}(16), 160502 (2007).

\bibitem{ref28}
H.-P.\ Breuer, E.-M.\ Laine, J.\ Piilo, and B.\ Vacchini, {Non-Markovian dynamics in open quantum systems}, Rev.\ Mod.\ Phys.\ {\bf 88}, 021002 (2016).

\bibitem{Ref27}
B.\ Bellomo, R.\ Lo~Franco, and G.\ Compagno, {Non-Markovian Effects on the dynamics of entanglement}, Phys.\ Rev.\ Lett.\ {\bf 99}, 160502 (2007).

\bibitem{Ref29}
W.\ Cui, Z.\ Xi, and Y.\ Pan, {Non-Markovian entanglement dynamics between two coupled qubits in the same environment}, J.\ Phys.\ A {\bf 42}, 155303 (2009).

\bibitem{Ref28}
C.\ S.\ Yu and H.\ S.\ Song, {Monogamy and entanglement in tripartite quantum states.}, Phys.\ Lett.\ A {\bf 373}(7), 727 (2009).

\bibitem{ref29}
H.\ P.\ Breuer and F.\ Petruccione, \textit{The Theory of Open Quantum Systems} (Oxford University Press, Oxford, UK, 2006).

\bibitem{r28}
F.\ F.\ Fanchini, M.\ F.\ Cornelio, M.\ C.\ De~Oliveira, \textit{et~al.}, {Conservation law for distributed entanglement of formation and quantum discord}, Phys.\ Rev.\ A {\bf 84}(1), 012313 (2010).

\bibitem{r29}
V.\ Madhok, V.\ Gupta, D.\ A.\ Trottier, \textit{et~al.}, {Signatures of chaos in the dynamics of quantum discord}, Phys.\ Rev.\ E {\bf 91}(3), 032906 (2015).

\bibitem{ref30}
G.\ Salton, R.\ B.\ Mann, and N.\ C.\ Menicucci, {Acceleration-assisted entanglement harvesting and rangefinding}, New J.\ Phys.\ {\bf 17}, 035001 (2015).

\bibitem{ref31}
N.\ D.\ Birrell and P.\ Davies, \textit{Quantum Fields in Curved Space} (Cambridge University Press, Cambridge, UK, 1982).

\bibitem{ref32}
Z.\ Wang, W.\ Wu, and J.\ Wang, {Steady-state entanglement and coherence of two coupled qubits in equilibrium and nonequilibrium environments}, Phys.\ Rev.\ A {\bf 99}, 042320 (2019).

\bibitem{Ref32}
Z.\ Zhang and J.\ Wang, {Landscape, kinetics, paths and statistics of curl flux, coherence, entanglement and energy transfer in non-equilibrium quantum systems}, New J.\ Phys.\ {\bf 17}, 043053 (2015).

\bibitem{Ref33}
Z.\ Zhang and J.\ Wang, {Shape, orientation and magnitude of the curl quantum flux, the coherence and the statistical correlations in energy transport at nonequilibrium steady state}, New J.\ Phys.\ {\bf 17}, 093021 (2015).

\bibitem{Ref34}
A.\ Satz, {Then again, how often does the Unruh--DeWitt detector click if we switch it carefully?}, Class.\ Quantum Grav.\ {\bf 24}, 1719 (2007).

\bibitem{ref33}
D.\ M.\ T.\ Benincasa, L.\ Borsten, M.\ Buck, and F.\ Dowker, {Quantum information processing and relativistic quantum fields}, Class.\ Quantum Grav.\ {\bf 31}, 075007 (2014).

\bibitem{ref35}
J.\ Marino, A.\ Noto, and R.\ Passante, {Thermal and nonthermal signatures of the Unruh effect in Casimir--Polder forces}, Phys.\ Rev.\ Lett.\ {\bf 113}, 020403 (2014).

\bibitem{ref34}
V.\ V.\ Dodonov, O.\ V.\ Man'ko, and V.\ I.\ Man'ko, {Photon distribution for one-mode mixed light with a generic Gaussian Wigner function}, Phys.\ Rev.\ A {\bf 49}, 2993 (1994).

\end{thebibliography}

\end{document}